%% file: Nf2_PFF.tex
\documentclass[11pt,prd,aps,floatfix,nofootinbib]{revtex4}

\usepackage{graphicx}
\usepackage{amsmath}
\usepackage{amssymb}
\usepackage{longtable}

\newcommand{\be}{\begin{equation}}
\newcommand{\ee}{\end{equation}}
\newcommand{\bea}{\begin{eqnarray}}
\newcommand{\eea}{\end{eqnarray}}
\newcommand{\bi}{\begin{itemize}}
\newcommand{\ei}{\end{itemize}}
\newcommand{\ben}{\begin{enumerate}}
\newcommand{\een}{\end{enumerate}}
\newcommand{\bt}{\begin{tabbing}}
\newcommand{\et}{\end{tabbing}}
\newcommand{\nn}{\nonumber}

\newcommand{\sgn}{{\rm sgn}}
\newcommand{\calO}{{\mathcal O}}
\newcommand{\Hw}{H_{\rm W}}

\newcommand{\bfp}{{\bf p}}
\newcommand{\bfq}{{\bf q}}
\newcommand{\bfk}{{\bf k}}
\newcommand{\bfx}{{\bf x}}
\newcommand{\bfr}{{\bf r}}
\newcommand{\crad}{\langle r^2 \rangle}

\begin{document}

\input{0.title}
\clearpage
\input{1.intro}

\input{2.simulation}
\input{3.pff}

\input{4.q2_interp}

\input{5.chiral_fit}

\input{6.conclusion}
\input{A.reference}

\end{document}

%% file: 0.title.tex
\vspace*{-10mm}
\begin{flushright}
\normalsize
KEK-CP-223        \\
\end{flushright}

\title{
Pion form factors from two-flavor lattice QCD with exact chiral symmetry
}

\author{
   S.~Aoki$^{a,b}$, 
   T.~W.~Chiu$^{c}$, 
   H.~Fukaya$^{d}$, 
   S.~Hashimoto$^{e,f}$, 
   T.~H.~Hsieh$^g$,
   T.~Kaneko$^{e,f}$, 
   H.~Matsufuru$^{h}$, 
   J.~Noaki$^{e}$, 
   T.~Onogi$^{i}$,  
   E.~Shintani$^{i}$
   and 
   N.~Yamada$^{e,f}$ \\
   (JLQCD and TWQCD collaborations)
}

\affiliation{
   $^a$Graduate School of Pure and Applied Sciences, 
   University of Tsukuba, Ibaraki 305-8571, Japan
   \\
   $^b$Riken BNL Research Center, 
   Brookhaven National Laboratory, Upton, New York 11973, USA
   \\
   $^c$Physics Department, Center for Theoretical Sciences,
   and Center for Quantum Science and Engineering,
   National Taiwan University, Taipei, 10617, Taiwan
   \\
   $^d$Department of Physics, Nagoya University, 
   Nagoya 464-8602, Japan
   \\
   $^e$KEK Theory Center, 
   High Energy Accelerator Research Organization (KEK),
   Ibaraki 305-0801, Japan 
   \\
   $^f$School of High Energy Accelerator Science,
   The Graduate University for Advanced Studies (Sokendai),
   Ibaraki 305-0801, Japan
   \\
   $^g$Research Center for Applied Sciences,
   Academia Sinica, Taipei 115, Taiwan
   \\
   $^h$Computing Research Center, 
   High Energy Accelerator Research Organization (KEK),
   Ibaraki 305-0801, Japan 
   \\ 
   $^i$Yukawa Institute for Theoretical Physics, Kyoto University,
   Kyoto 606-8502, Japan
}

\date{\today}

\begin{abstract}

We calculate pion vector and scalar form factors 
in two-flavor lattice QCD and study the chiral behavior 
of the vector and scalar radii $\crad_{V,S}$.
Numerical simulations are carried out on a $16^3 \times 32$
lattice at a lattice spacing of 0.12 fm with quark masses down to 
$\sim \! m_s/6$,
where $m_s$ is the physical strange quark mass. 
Chiral symmetry,
which is essential for a direct comparison with 
chiral perturbation theory (ChPT),
is exactly preserved in our calculation 
at finite lattice spacing by employing the overlap quark action.
We utilize the so-called all-to-all quark propagator
in order to calculate
the scalar form factor including the contributions of disconnected diagrams
and to improve statistical accuracy of the form factors.
A detailed comparison with ChPT reveals that 
the next-to-next-to-leading-order contributions to the radii 
are essential to describe their chiral behavior 
in the region of quark mass from $m_s/6$ to $m_s/2$.
Chiral extrapolation based on two-loop ChPT 
yields $\crad_V \!=\! 0.409(23)(37)~\mbox{fm}^2$ and 
$\crad_S \!=\! 0.617(79)(66)~\mbox{fm}^2$, 
which are consistent with phenomenological analysis.
We also present our estimates of relevant low-energy constants.

\end{abstract}


\maketitle

%% file: 1.intro.tex

\section{Introduction}


Recent algorithmic improvements allow us to perform 
large-scale simulations of unquenched QCD in the chiral regime 
with various lattice discretizations.
Calculation of phenomenologically important quantities has then become feasible.
In order to make reliable prediction for physical observables,
it is crucial to 
examine the consistency of their chiral behavior 
with expectations from chiral perturbation theory (ChPT).
Lattice QCD with exact chiral symmetry 
is the cleanest framework for this purpose,
while conventional lattice actions
may distort chiral behavior of observables by their explicit symmetry breaking, 
especially when one goes beyond the next-to-leading order (NLO) in ChPT.
The JLQCD and TWQCD collaborations embarked on 
the simulations with exact chiral symmetry
\cite{Lat07:JLQCD:Matsufuru,Lat08:JLQCD:Hashimoto}
employing the overlap quark action \cite{Overlap:NN,Overlap:N}.
So far, we have performed a detailed study of 
the pion mass and decay constant in two-flavor lattice QCD
\cite{Spectrum:Nf2:RG+Ovr:JLQCD}.
For other physics results from this project, see \cite{e-regime:eigen:Nf2:RG+Ovr:JLQCD:1,chi_t:Nf2:RG+Ovr:JLQCD+TWQCD,e-regime:msn:Nf2:RG+Ovr:JLQCD,e-regime:eigen:Nf2:RG+Ovr:JLQCD:2,B_K:Nf2:RG+Ovr:JLQCD,S:Nf2:RG+Ovr:JLQCD,Sigma:Nf2:RG+Ovr:JLQCD,OPE:Nf2:RG+Ovr:JLQCD}.


The pion vector form factor $F_V(q^2)$ defined by
\bea
   \langle \pi(p^\prime) | V_\mu | \pi(p) \rangle 
   & = &
   \left( p + p^\prime \right)_\mu F_V(q^2),
   \hspace{5mm}
   q^2=(p-p^\prime)^2,
   \label{eqn:intro:pff_v}
\eea
provides a simple testing ground for the consistency 
between lattice calculation and ChPT.
This is one of the fundamental quantities 
to characterize the low-energy dynamics of pions:
for instance, it is related to the charge radius of pion
\bea
   \crad_V 
   & = & 
   6 \left. \frac{\partial F_V(q^2)}{\partial q^2} \right|_{q^2=0}.
\eea
The chiral expansion of $F_V(q^2)$ and $\crad_V$ is available up to two loops, 
namely to the next-to-next-to-leading-order (NNLO), 
both for $N_f\!=\!2$ \cite{PFF:ChPT:NLO:1,PFF_V:ChPT:NNLO,PFF_V:ChPT:NNLO:2}
and $N_f\!=\!3$ \cite{PFF:ChPT:NLO:2,PFF_V:ChPT:NNLO:Nf3}.
Analyses of experimental data based on two-loop ChPT 
have led to precise estimates of $\crad_V$ 
\cite{PFF_V:ChPT:NNLO,PFF_V:ChPT:NNLO:Nf3},
which can also be used as a benchmark of lattice calculations.

From previous lattice studies 
\cite{PFF:Nf2:Plq+Clv:HKL,PFF:Nf2:DBW2+DW:RBC,PFF:Nf0:Plq+tmW:AL,PFF:impG+AT+DWF:LHP,PFF:Nf2:Plq+Clv:JLQCD,PFF:Nf2:Plq+Clv:QCDSF,PFF:Nf0:LW+CImp:BGR,PFF:Nf3:RG+DW:RBC+UKQCD,PFF:Nf2:Sym+tmW:ETMC}, 
the consistency with experiment has not been established convincingly:
some of them reported good agreement with experiment
\cite{PFF:Nf2:Plq+Clv:QCDSF,PFF:Nf3:RG+DW:RBC+UKQCD,PFF:Nf2:Sym+tmW:ETMC},
whereas others underestimated $\crad_V$ significantly
\cite{PFF:Nf2:Plq+Clv:HKL,PFF:impG+AT+DWF:LHP,PFF:Nf2:Plq+Clv:JLQCD,PFF:Nf0:LW+CImp:BGR}.
This is possibly due to systematics 
in the parametrization of the $q^2$ dependence of $F_V(q^2)$ 
and to the chiral extrapolation of $\crad_V$.
For instance, 
the NNLO chiral corrections are fully taken into account in the ChPT analyses 
but not in most of the previous lattice studies.
They could significantly modify the chiral behavior of $\crad_V$ 
at up and down quark masses larger than their physical value,
as demonstrated in our report~\cite{Lat08:JLQCD:TK}
and more recently in Ref.~\cite{PFF:Nf2:Sym+tmW:ETMC}
with a different lattice discretization.
Our simulations with exact chiral symmetry enable us 
to perform more stringent test of the chiral behavior of lattice data 
using two-loop ChPT
without suffering from the distortion due to explicit chiral symmetry breaking
present in other frameworks.


The chiral behavior of the scalar form factor $F_S(q^2)$ defined by 
\bea
   \langle \pi(p^\prime) | S | \pi(p) \rangle 
   & = &
   F_S(q^2) 
   \label{eqn:intro:pff_s}
\eea
is another interesting subject, since 
i) its radius 
\bea
   \crad_S
   & = & 
   6 \left. \frac{\partial F_S(q^2)}{\partial q^2} \right|_{q^2=0}
\eea
provides a determination of the low energy constant (LEC) $l_4$ 
alternative to that with the decay constant $F_\pi$, and 
ii) $\crad_S$ has 6~times enhanced chiral logarithm compared to $\crad_V$
and thus may offer an opportunity to clearly identify
the one-loop chiral logarithm.
Since there are no experimental processes directly related to $F_S(q^2)$,
its direct determination is possible only through lattice QCD.
It is however difficult to evaluate 
disconnected correlation functions 
with conventional point-to-all quark propagator, 
which flows from a fixed lattice site to any site.
There have been only a few calculations of $F_S(q^2)$
\cite{PFF:Nf2:Plq+Clv:JLQCD,PFF:Nf0:LW+CImp:BGR},
and the contributions of disconnected diagrams were ignored in these studies.


In this paper,
we calculate the pion vector and scalar form factors
$F_{V,S}(q^2)$ in two-flavor QCD 
and study the chiral behavior of the radii $\crad_{V,S}$.
For a detailed comparison with two-loop ChPT, 
we preserve chiral symmetry by employing the overlap quark action 
and simulate up and down quarks with masses as low as $m \! \sim \! m_s/6$.
The scalar form factor $F_S(q^2)$ is evaluated including the contribution of 
disconnected diagrams by using the all-to-all quark propagator,
which contains propagations from {\it any} lattice site to any site. 
The all-to-all propagator is also helpful to substantially improve 
statistical accuracy of $F_{V,S}(q^2)$.
Our preliminary analyses based on one- and two-loop ChPT 
have been reported in Refs.~\cite{Lat07:JLQCD:TK} and \cite{Lat08:JLQCD:TK},
respectively.


This paper is organized as follows.
We introduce our simulation method in Sec.~\ref{sec:sim}.
Calculation of the form factors from pion correlators is 
presented in Sec~\ref{sec:pff}.
We parametrize their $q^2$ dependence in Sec.~\ref{sec:q2_interp}. 
Section~\ref{sec:chiral_fit} is devoted to a detailed description 
of our chiral extrapolation of the radii.
Finally, our concluding remarks are given in Sec.~\ref{sec:concl}.

%% file: 2.simulation.tex

\section{Simulation method}
\label{sec:sim}

\subsection{Configuration generation}


We calculate pion form factors in QCD with dynamical up and down quarks
with a degenerate mass parameter.
Numerical simulations are carried out 
with the Iwasaki gauge action \cite{Iwasaki} 
and the overlap quark action \cite{Overlap:NN,Overlap:N}, 
which has exact chiral symmetry at finite lattice spacings
\cite{lat_chial_sym}.
Its Dirac operator is given by 
\bea
   D(m)
   & = &
   \left( m_0 + \frac{m}{2} \right)
   +
   \left( m_0 - \frac{m}{2} \right)  
   \gamma_5 \, \sgn \left[ \Hw (-m_0) \right],
   \label{eqn:sim:conf_gen:overlap}
\eea   
where $m$ is the quark mass and 
$\Hw(-m_0) \!=\! \gamma_5 D_{\rm W}(-m_0)$ 
is the Hermitian Wilson-Dirac operator.
We set the mass parameter of this kernel operator to $-m_0\!=\!-1.6$, 
with which the locality of the overlap Dirac operator is confirmed
\cite{Lat06:JLQCD:Yamada,Prod_Run:JLQCD:Nf2:RG+Ovr}.
%
%
Because of the sign function $\sgn[\Hw(-m_0)]$ in $D(m)$,
the overlap action is discontinuous when $\Hw(-m_0)$ 
develops zero eigenvalue(s).
The commonly-used Hybrid Monte Carlo (HMC) algorithm 
can be modified to deal with this discontinuity  \cite{overlap:HMC:FKS} 
but turned out to be very costly.
In order to carry out high-statistics simulations, 
we suppress (near-)zero modes of $\Hw(-m_0)$
by introducing an auxiliary determinant 
\bea
   \Delta_{\rm W} 
   & = & 
   \frac{\det[\Hw(-m_0)^2]}{\det[\Hw(-m_0)^2+\mu^2]}
   \label{eqn:sim:conf_gen:det}
\eea
into the Boltzmann weight \cite{exW:Vranas,exW+extmW:JLQCD}.
We note that this can be considered as an $O(a^2)$ modification of 
the gauge action,
and hence does not change the continuum limit of the theory.
The parameter $\mu$ is tuned to 0.2 
so as to minimize lattice artifacts 
induced by $\Delta_{\rm W}$
\cite{Lat06:JLQCD:Yamada,exW+extmW:JLQCD}.
%
%
We refer readers to Ref.~\cite {Prod_Run:JLQCD:Nf2:RG+Ovr}
for further details of our simulation method.


An important property of the determinant $\Delta_{\rm W}$ is that 
it fixes the global topological charge $Q$ of the gauge field 
during continuous updating of the gauge configuration in the HMC algorithm.
Note, however, that
local topological fluctuations are present,
and the topological susceptibility calculated 
in Ref.~\cite{chi_t:Nf2:RG+Ovr:JLQCD+TWQCD} shows expected behavior 
as a function of sea quark mass.
As shown in Refs.~\cite{fixed_Q:BCNW,fixed_Q:AFHO},
the effect of fixed {\it global} topology can be considered 
as a finite volume effect
and is suppressed by the inverse of the space-time volume $V$.
Furthermore the effect can be systematically corrected 
by investigating $Q$ dependence of physical observables of interest
~\cite{fixed_Q:AFHO}.


Our gauge ensembles are generated on a $16^3 \times 32$ lattice.
The bare gauge coupling is $\beta\!=6/g^2\!=\!2.30$
where the lattice spacing 
determined from the Sommer scale $r_0\!=\!0.49$\,fm \cite{r0}
is $a\!=\!0.1184(21)$\,fm.
In the trivial topological sector $Q\!=\!0$,
we take four values of bare up and down quark masses,
$m\!=\!0.015$, 0.025, 0.035 and 0.050,
which cover a range $[m_s/6,m_s/2]$.
At each $m$,
we calculate pion form factors using 100 independent configurations 
separated by 100 HMC trajectories.
In order to study the effect of fixed topology,
we also simulate non-trivial topological sectors $Q\!=\!-2$ and $-4$
at $m\!=\!0.050$ 
with statistics of 50 independent configurations.

\subsection{Construction of all-to-all quark propagator}

The conventional method to calculate hadron correlators
employs the so-called point-to-all quark propagator
which flows from a fixed lattice site to any site.
This is however not suitable to calculate disconnected diagrams
which involve quark loops starting from and ending at arbitrary lattice sites.
In this work, therefore,
we construct all-to-all quark propagator 
that contains the quark propagating from any lattice site to any site 
along the strategy proposed in Ref.~\cite{A2A}.

Let us consider a decomposition of the quark propagator 
using eigenmodes of the overlap operator
\bea
   D^{-1}(x,y)
   & = &
   \sum_k \frac{1}{\lambda^{(k)}} u^{(k)}(x) u^{(k)\dagger}(y),
\eea
where $\lambda^{(k)}$ and $u^{(k)}(x)$ represent $k$-th lowest eigenvalue 
and eigenvector of $D$, respectively.
Note that the eigenvalues are ordered by their absolute values.
Color and spinor indices are suppressed for simplicity.
It is expected that low-lying modes dominate
low-energy dynamics of pions including their form factors.
We evaluate these low-mode contributions to the propagator 
{\it exactly} as
\bea
   (D^{-1})_{\rm low}(x,y)
   & = &
   \sum_{k=1}^{N_e}
      \frac{1}{\lambda^{(k)}}\,u^{(k)}(x)u^{(k)\dagger}(y)
   \hspace{5mm}
   (N_e\!=\!100)
   \label{eqn:sim:a2a_prop:low}
\eea
using 100 low-lying modes for each gauge configuration.
Note that the overlap operator is normal and 
we do not have to distinguish left and right eigenvectors.

The contribution of higher modes is estimated stochastically 
by the noise method with the dilution technique \cite{A2A}.
We prepare a single $Z_2$ noise vector $\eta^{(d)}(x)$ for each configuration,
and {\it dilute} it into
\noindent
$N_d = 3 \times 4 \times N_t/2$ vectors,
which have non-zero elements only for a single combination of 
color and spinor indices and at two consecutive time-slices.
The high-mode contribution can be estimated as 
\bea
   (D^{-1})_{\rm high}(x,y)
   & = & 
   \sum_{d=1}^{N_d} x^{(d)}(x)\,\eta^{(d)\dagger}(y)
   \label{eqn:sim:a2a_prop:high}
\eea
by solving a linear equation for each diluted source
\bea
   \sum_y
   D(x,y)\,x^{(d)}(y)
   = 
   \sum_y
   (\delta_{xy}-P_{\rm low}(x,y)) \, \eta^{(d)}(y)
   \hspace{5mm}           
   (d=1,...,N_d),
   \hspace{2mm}           
   \label{eqn:sim:a2a_prop:high:leq}
\eea
where $d$ is an index to represent the dilution and 
\bea
   P_{\rm low}(x,y)
   & = & 
   \sum_{k=1}^{N_e} u^{(k)}(x)\, u^{(k) \dagger}(y)
   \label{eqn:sim:a2a_prop:proj:low}
\eea
is the projector to the eigenspace spanned by the low-modes. 

By combining 
Eqs.~(\ref{eqn:sim:a2a_prop:low}) and (\ref{eqn:sim:a2a_prop:high}),
the all-to-all quark propagator can be expressed by a matrix
\bea
   D^{-1}(x,y)
   & = &
   \sum_{k=1}^{N_v} v^{(k)}(x)\,w^{(k)\dagger}(y)
   \label{eqn:sim:a2a_prop:a2a_prop}
\eea
constructed from the following two sets of vectors $v$ and $w$
\bea
   \left\{
      v^{(1)},...,v^{(N_v)}
   \right\}
   & = & 
   \left\{
      \frac{u^{(1)}}{\lambda^{(1)}},
      \ldots,
      \frac{u^{(N_e)}}{\lambda^{(N_e)}},
      x^{(1)}, \dots, x^{(N_d)}
   \right\},
   \label{eqn:sim:a2a_prop:a2a:vw_vectors:v}
   \\
   \left\{
      w^{(1)},...,w^{(N_v)}
   \right\}
   & = &
   \left\{
      u^{(1)}, \ldots, u^{(N_e)},
      \eta^{(1)}, \dots, \eta^{(N_d)}
   \right\},
   \label{eqn:sim:a2a_prop:a2a:vw_vectors:w}
\eea
where $N_v = N_e+N_d$.

\subsection{Measurement of pion correlators}


Using the all-to-all propagator (\ref{eqn:sim:a2a_prop:a2a_prop}),
the pion two-point function  
with a temporal separation $\Delta t$ and a spatial momentum $\bfp$
can be expressed as 
\bea
    C_{\pi \pi, \phi \phi^\prime}(\Delta t; \bfp)
    & =  &
    \frac{1}{N_t}
    \sum_{t=1}^{N_t}
    \sum_{k,l=1}^{N_v}
    \calO^{(k,l)}_{\gamma_5,\phi^\prime}(t+\Delta t,\bfp) \, 
    \calO^{(l,k)}_{\gamma_5,\phi}(t,-\bfp).
    \label{eqn:sim:corr:msn_corr_2pt} 
\eea
Here $\calO_{\Gamma,\phi}(t,\bfp)$ is constructed 
from the $v$ and $w$ vectors as 
\bea
   \calO^{(k,l)}_{\Gamma,\phi}(t;\bfp)
   & = & 
   \sum_{\bfx,\bfr}
   \phi(\bfr)\, 
   w^{(k)\dagger}(\bfx+\bfr,t) \, 
   \Gamma \,
   v^{(l)}(\bfx,t)\,
   e^{-i \bfp \bfx},
   \label{eqn:sim:corr:msn_op} 
\eea
and represents the meson field 
with a Dirac matrix $\Gamma$ and a momentum $\bfp$ 
at a temporal coordinate $t$.
We use a local $\phi_{l}(\bfr) \!=\! \delta_{\bfr,{\bf 0}}$
and an exponential form $\phi_{s}(\bfr) \!=\! \exp[-0.4|\bfr|]$
for the smearing function $\phi(\bfr)$ in this study.
Note that the source point $\bfx$ is averaged over spatial volume.

\begin{figure}[t]
\begin{center}
\includegraphics[angle=0,width=0.3\linewidth,clip]%
                {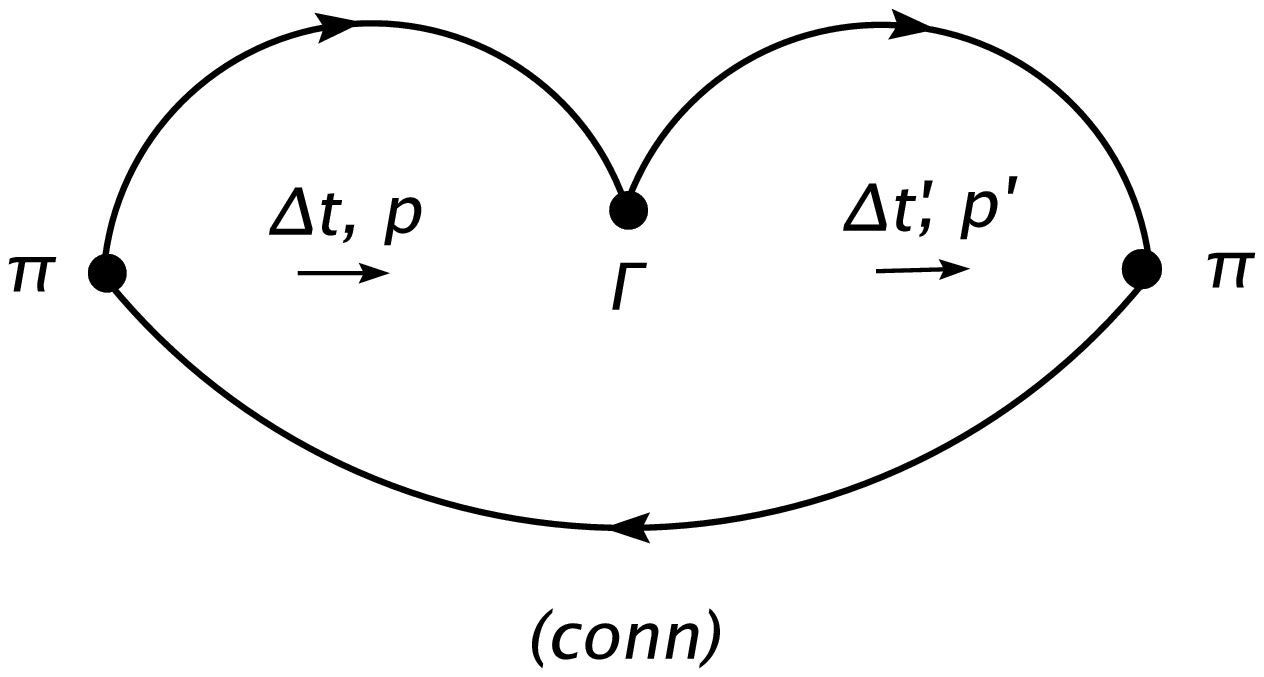}
\hspace{5mm}
\includegraphics[angle=0,width=0.3\linewidth,clip]%
                {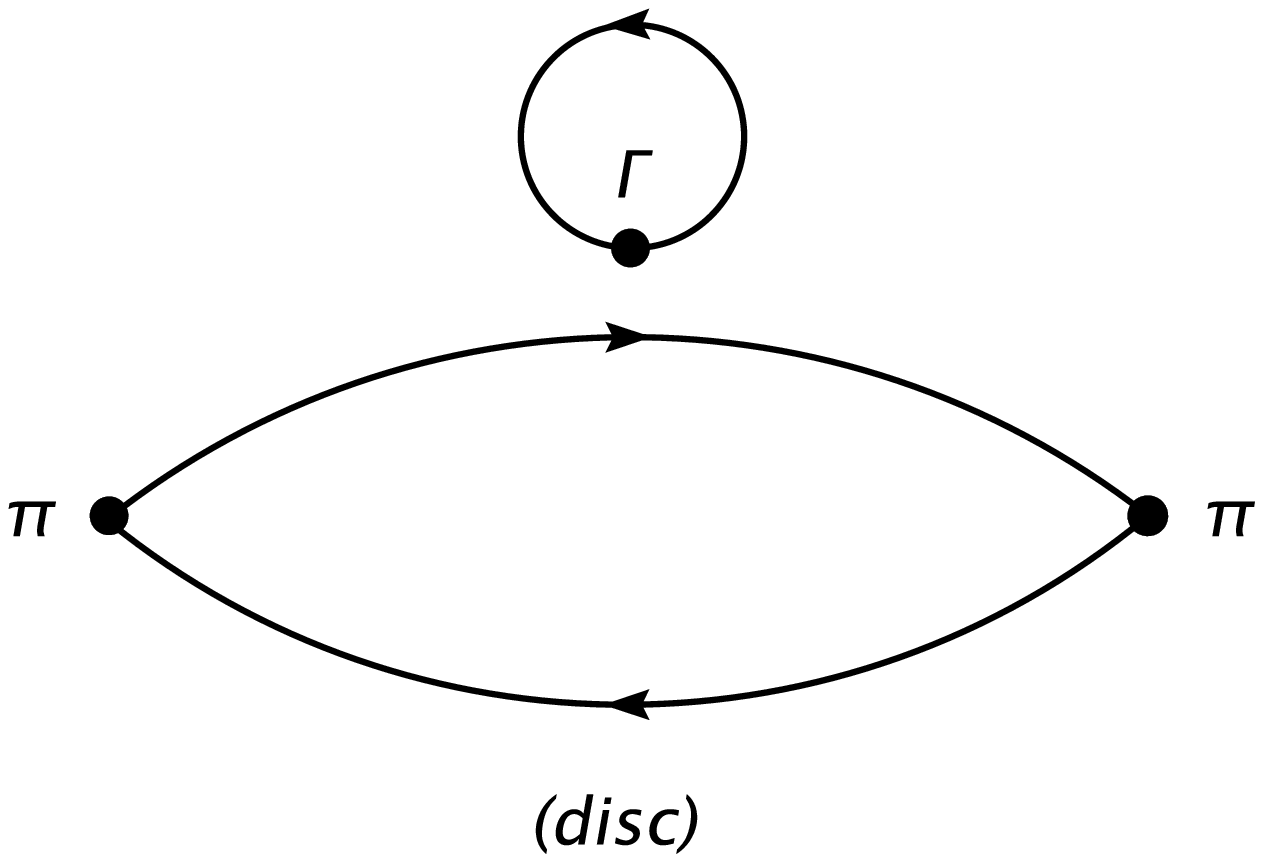}
\hspace{5mm}
\includegraphics[angle=0,width=0.3\linewidth,clip]%
                {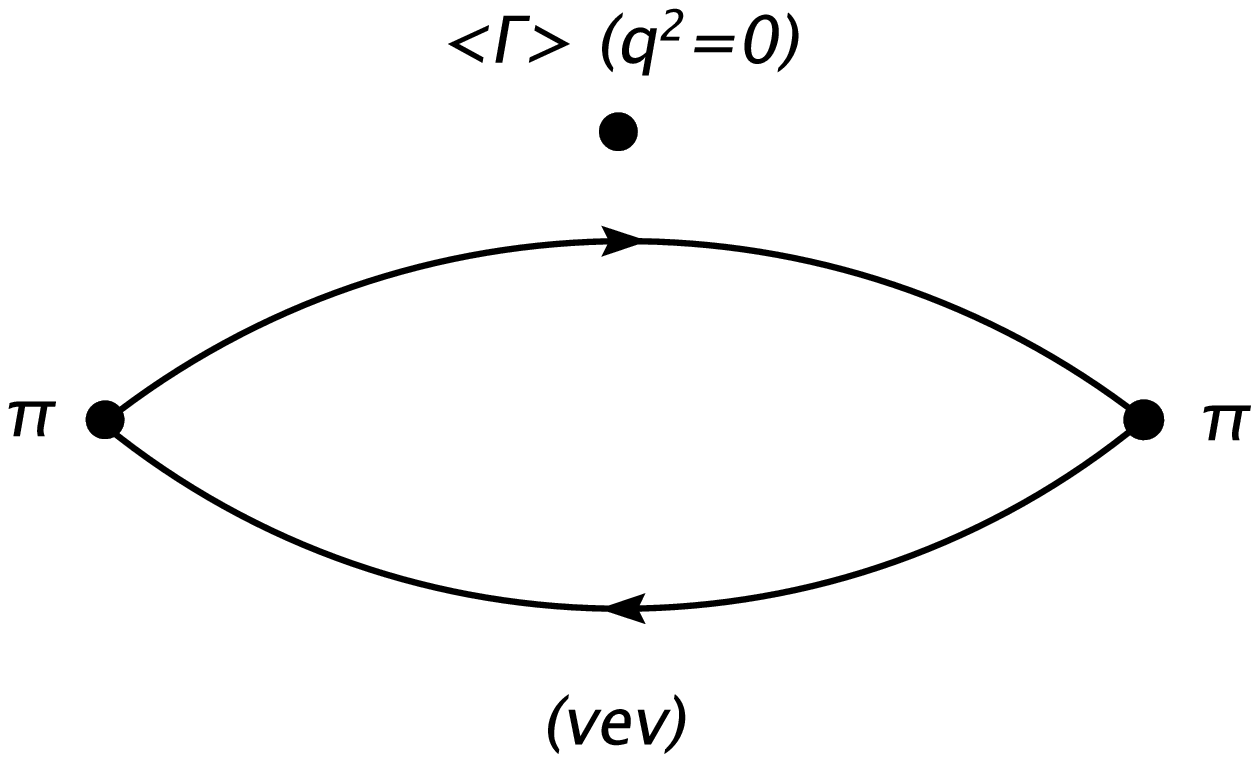}
\vspace{-2mm}
\caption{
   Connected (left-most diagram) and disconnected three point functions 
   (middle diagram).
   Note that $F_S(0)$ receives 
   a contribution from the right-most diagram 
   due to the non-zero vacuum expectation value (VEV)
   of the scalar operator $S$.
}
\label{fig:sim:corr:msn_corr_3pt:diag}
\end{center}
\end{figure}

Pion form factors are extracted from three-point functions 
shown in Fig.~\ref{fig:sim:corr:msn_corr_3pt:diag}, 
which can also be calculated from the meson fields
$\calO_{\Gamma,\phi}(t,\bfp)$ as 
\bea
    C_{\pi \Gamma \pi}^{\rm (conn)}
    (\Delta t, \Delta t^{\prime};\bfp,\bfp^\prime)
    & = & 
    \frac{1}{N_t} \sum_{t=1}^{N_t}
    \sum_{k,l,m=1}^{N_v}
    \calO^{(m,l)}_{\gamma_5,\phi_s}(t+\Delta t + \Delta t^\prime;\bfp^\prime) \, 
    \calO^{(l,k)}_{\Gamma,\phi_l}(t+ \Delta t;\bfp-\bfp^\prime)
    \nn \\[-2mm]
    & &
    \hspace{50mm}
    \times
    \calO^{(k,m)}_{\gamma_5,\phi_s}(t;-\bfp),
    \label{eqn:sim:corr:msn_corr_3pt:conn}  
\eea
\bea
    C_{\pi \Gamma \pi}^{\rm (disc)}
    (\Delta t, \Delta t^\prime;\bfp,\bfp^\prime)
    & = & 
    \frac{1}{N_t} \sum_{t=1}^{N_t}
    \sum_{k,l=1}^{N_v}
    \calO^{(k,l)}_{\gamma_5,\phi_s}(t + \Delta t + \Delta t^\prime;\bfp^\prime) \, 
    \calO^{(l,k)}_{\gamma_5,\phi_s}(t;-\bfp)
    \nn \\[-2mm]
    & &
    \hspace{50mm}
    \times
    \sum_{k=1}^{N_v}
    \calO^{(k,k)}_{\Gamma,\phi_{l}}(t+\Delta t;\bfp-\bfp^\prime), 
    \label{eqn:sim:corr:msn_corr_3pt:disc}  
    \\
    C_{\pi \Gamma \pi}^{\rm (vev)}
    (\Delta t, \Delta t^\prime;\bfp,\bfp^\prime)
    & = & 
    \frac{1}{N_t} \sum_{t=1}^{N_t}
    \sum_{k,l=1}^{N_v}
    \calO^{(k,l)}_{\gamma_5,\phi_s}(t+\Delta t + \Delta t^\prime;\bfp^\prime) \, 
    \calO^{(l,k)}_{\gamma_5,\phi_s}(t;-\bfp)
    \nn \\[-2mm]
    &&
    \hspace{35mm}
    \times 
    \left\langle
       \frac{1}{N_t} \sum_{t^\prime=1}^{N_t}
       \sum_{k=1}^{N_v}
       \calO^{(k,k)}_{\Gamma,\phi_{l}}(t^\prime;\bfp-\bfp^\prime)
    \right\rangle_{\rm conf},
    \label{eqn:sim:corr:msn_corr_3pt:vev}  
\eea
where $\langle \cdots \rangle_{\rm conf}$ represents the ensemble average.
We denote the temporal separation and spatial momentum for the initial 
(final) meson by $\Delta t$ and $\bfp$ ($\Delta t^\prime$ and $\bfp^\prime$),
respectively.
%


We prepare the $v$ and $w$ vectors on the IBM Blue Gene/L at KEK,
which has the peak speed of 57.3~TFLOPS.
We employ the implicitly restarted Lanczos algorithm
to calculate low-modes of $D$. 
The computational cost of this step is roughly
$0.6~\mbox{TFLOPS} \cdot \mbox{hours}$ per configuration.
Solving Eq.~(\ref{eqn:sim:a2a_prop:high:leq}) is the most time-consuming part 
in our measurement, 
since it requires $N_t/2$ times more inversions 
than the conventional measurement of two-point functions 
with the point-to-all propagator.
We use the conjugate gradient (CG) algorithm
accelerated by the relaxed stopping condition \cite{relCG}
and by the preconditioning using the 100 low-modes.
The resulting CPU cost of our CG solver is 
$\sim 1.7~\mbox{TFLOPS} \cdot \mbox{hours} / \mbox{conf}$.
The calculation of the meson field $\calO^{(k,l)}_{\Gamma,\phi}(t,\bfp)$
needs much less CPU time than the above two steps:
it is about $0.2~\mbox{GFLOPS} \cdot \mbox{hours} / \mbox{conf}$
for a single choice of $(m,\bfp,\Gamma,\phi)$.
Once $\calO^{(k,l)}_{\Gamma,\phi}(t,\bfp)$ is prepared, 
we can calculate all of the connected and disconnected pion correlators 
with small additional cost.
These calculations are carried out on the Hitachi SR11000 
with the peak speed of 2.15~TFLOPS
and workstations at KEK.


It is advantageous that
we do not have to repeat  the time consuming preparation of 
$v$ and $w$ vectors
to calculate meson fields $\calO_{\Gamma,\phi}(t,\bfp)$ 
with different choices of $\bfp$, $\Gamma$ and $\phi$.
In order to simulate various values of $q^2$,
we take 27 choices of the spatial momentum $\bfp$ with $|\bfp|\!\leq\!\sqrt{3}$
for the initial and final pions, and 
33 choices with $|\bfq|\!\leq\!2$ for the momentum transfer $q$.
Note that the spatial momentum is shown in units of $2\pi/L$ in this article.
This setup covers the region of momentum transfer
$-1.7 \! \lesssim q^2~\mbox{[GeV$^2$]} \! \leq \! 0$.

\begin{figure}[t]
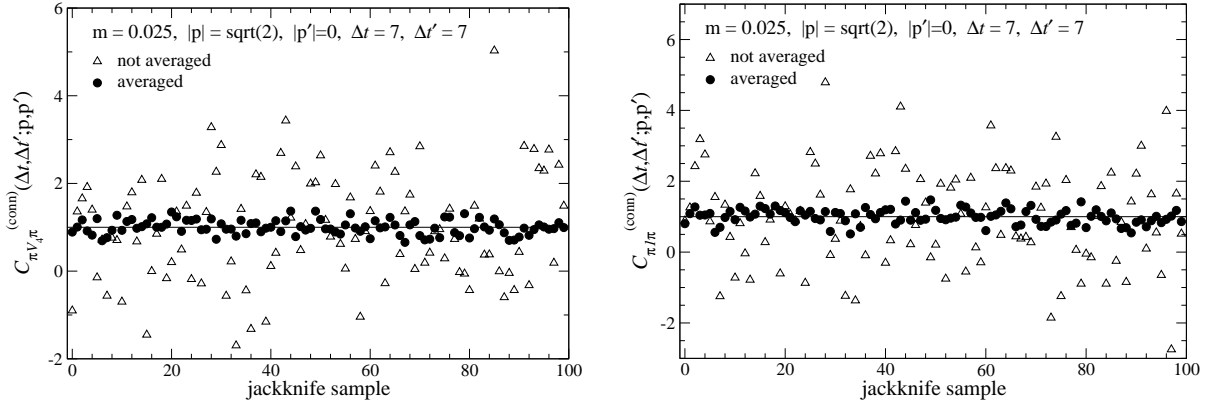

\begin{center}
   \includegraphics[angle=0,width=0.48\linewidth,clip]{msn_3pt_p-p-v4_mud1_mval111_mom0700_smr11.eps}
   \hspace{3mm}
   \includegraphics[angle=0,width=0.48\linewidth,clip]{msn_3pt_p-p-s_mud1_mval111_mom0700_smr11.eps}

   \vspace{-2mm}
   \caption{
      Statistical fluctuation of connected three point functions
      $C_{\pi V_4 \pi, \phi_s \phi_s}^{{\rm (conn)}}(\Delta t,\Delta t^\prime;\bfp,\bfp^\prime)$
      (left panel) and 
      $C_{\pi 1 \pi, \phi_s \phi_s}^{{\rm (conn)}}(\Delta t,\Delta t^\prime;\bfp,\bfp^\prime)$
      (right panel) with $\Delta t\!=\!\Delta t^\prime\!=\!7$,
      $|\bfp|\!=\!\sqrt{2}$ and $|\bfp^\prime|\!=\!0$.
      Open triangles are jackknife data 
      with a single choice of the source location $(\bfx,t)$ and 
      momentum configuration $(\bfp,\bfp^\prime)$,
      whereas the filled squares are averaged over 
      $(\bfx,t)$ and $(\bfp,\bfp^\prime)$ 
      corresponding to the same value of $q^2$.
      Each data is normalized by its statistical average.     
   }
   \label{fig:sim:a2a:corr:jkd}     
\end{center}
\end{figure}


The use of the all-to-all propagator also 
improves the statistical accuracy of the pion correlators
by averaging over the locations of source operator\footnote{
We note in passing that
only the low-mode contribution 
is averaged over the locations of the source operator 
in the so-called low-mode averaging (LMA) method \cite{LMA:1,LMA:2}.
The high-mode contribution in LMA is estimated 
for a fixed source location using the point-to-all propagator
in contrast to our method with the all-to-all propagator. 
},
namely $(\bfx,t)$ in Eqs.~(\ref{eqn:sim:corr:msn_op})\,--\,(\ref{eqn:sim:corr:msn_corr_3pt:vev}),
as well as over momentum configurations $(\bfp,\bfp^\prime)$
corresponding to the same value of $q^2$.
Figure~\ref{fig:sim:a2a:corr:jkd} compares 
the statistical fluctuation of connected pion correlators 
with a certain choice of $\Delta t$, $\Delta t^\prime$ and $q^2$.
We observe that 
averaging over the source locations and momentum configurations
remarkably reduces the statistical error of the correlators
and leads to very accurate results for the form factors.


%

%% file: 3.pff.tex
\section{Determination of pion form factors}
\label{sec:pff}


\subsection{Ratio method for vector form factor}

In the limit of large temporal separations among pion operators and 
vector current,
two- and three-point functions are dominated 
by the contribution from the ground state
\bea
   C_{\pi\pi,\phi\phi^\prime}(\Delta t;\bfp)
   & \xrightarrow[\Delta t \to \infty]{} &
   \frac{Z_{\pi,\phi^\prime}(|\bfp|)^*\, Z_{\pi, \phi}(|\bfp|)} 
        {2\, E_\pi(|\bfp|)}\,
   e^{-E_\pi(|\bfp|)\,\Delta t},
   \label{eqn:pff:pff_v:msn_corr_2pt}
   \\
   C_{\pi \Gamma \pi}^{\rm (conn)}
   (\Delta t, \Delta t^{\prime};\bfp,\bfp^\prime)
   & \xrightarrow[\Delta t,\Delta t^\prime \to \infty]{} &
   \frac{Z_{\pi,\phi^\prime}(|\bfp^\prime|)^*\, Z_{\pi, \phi}(|\bfp|)} 
   {4\, E_\pi(|\bfp^\prime|) E_\pi(|\bfp|)}
   \frac{1}{Z_V}
   \langle \pi(p^\prime) | V_\mu | \pi(p) \rangle\,
   \nn \\
   &   &
   \hspace{40mm} 
   \times 
   e^{-E_\pi(|\bfp^\prime|)\, \Delta t^\prime}
   e^{-E_\pi(|\bfp|)\,        \Delta t},
   \label{eqn:pff:pff_v:msn_corr_3pt}
\eea
where $Z_{\pi,\phi}(|\bfp|)\!=\!\langle \pi(p) | \calO_{\gamma_5,\phi} \rangle$
is the overlap of the interpolating field $\calO_{\gamma_5,\phi}$ 
to the physical state, 
and $Z_V$ is the renormalization factor for the vector current. 
For a precise determination of the matrix element 
$\langle \pi(p^\prime) | V_\mu | \pi(p) \rangle$,
it is advantageous to take a ratio of appropriately chosen correlators
in order to cancel out the exponential damping factors 
$e^{-E_\pi(|\bfp^{(\prime)}|)\Delta t^{(\prime)}}$
and other unnecessary factors $Z_{\pi,\phi^{(\prime)}}$ and $Z_V$
\cite{dble_ratio}.
In this study, 
we use the following ratio 
to calculate an effective value of the vector form factor $F_V(q^2)$ 
\bea
   F_V(\Delta t,\Delta t^\prime;q^2)
   & = &  
   \frac{2\,M_\pi}{E_\pi(|\bfp|)+E_\pi(|\bfp^\prime|)}
   \frac{R_V(\Delta t,\Delta t^\prime; |\bfp|,|\bfp^\prime|,q^2)}
        {R_V(\Delta t,\Delta t^\prime; 0,0,0)},
   \label{eqn:pff:pff_v:dratio}
   \\
   R_V(\Delta t,\Delta t^\prime; |\bfp|,|\bfp^\prime|,q^2)
   & = &
   \frac{1}{N_{|\bfp|,|\bfp^\prime|}} 
   \sum_{\mbox{\scriptsize fixed }|\bfp|,|\bfp^\prime|}
   \frac{C_{\pi \gamma_4 \pi}^{\rm (conn)}
         (\Delta t,\Delta t^\prime; \bfp,\bfp^\prime)}
        {C_{\pi \pi, \phi_s \phi_l}(\Delta t;\bfp)\,
         C_{\pi \pi, \phi_l \phi_s}(\Delta t^\prime;\bfp^\prime)},
   \label{eqn:pff:pff_v:ratio}
\eea       
where $(1/N_{|\bfp|,|\bfp^\prime|})\sum_{\mbox{\scriptsize fixed }|\bfp|,|\bfp^\prime|}$
represents the average over momentum configurations 
corresponding to the same value of $q^2$.

\begin{figure}[t]
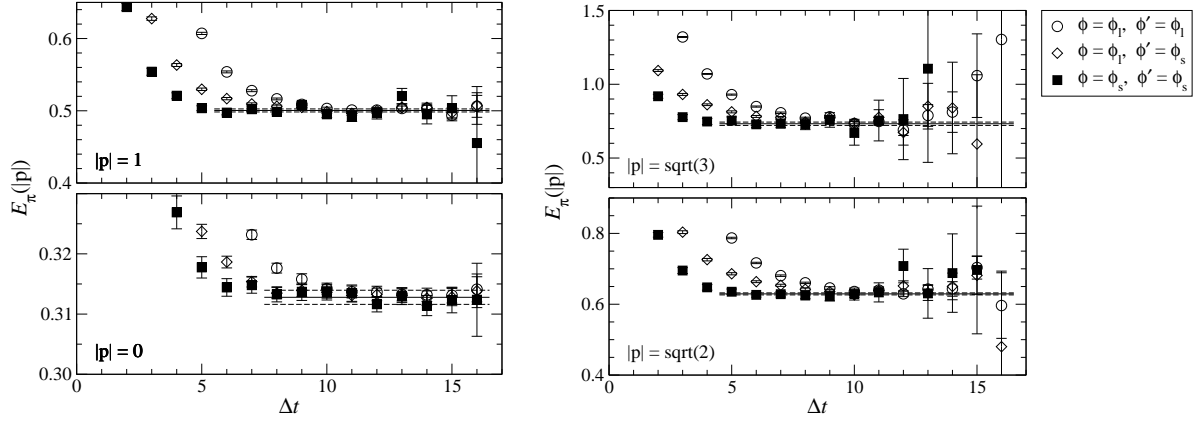

\begin{center}
\includegraphics[angle=0,width=0.41\linewidth,clip]%
                {em_msn_pp_mud3_mval33_tsrc00_set1.eps}
\hspace{3mm}
\includegraphics[angle=0,width=0.54\linewidth,clip]%
                {em_msn_pp_mud3_mval33_tsrc00_set2.eps}

\vspace{-3mm}
\caption{
   Effective value of pion energy $E_\pi(|\bfp|)$ at $(Q,m)\!=\!(0,0.050)$.
   Circles, triangles and squares show results from 
   $C_{\pi\pi,\phi_l \phi_l}(\Delta t;\bfp)$,
   $C_{\pi\pi,\phi_l \phi_s}(\Delta t;\bfp)$ and 
   $C_{\pi\pi,\phi_s \phi_s}(\Delta t;\bfp)$, respectively.
}
\label{fig:pff:Epi:m0050}
\end{center}
\end{figure}

\begin{figure}[t]
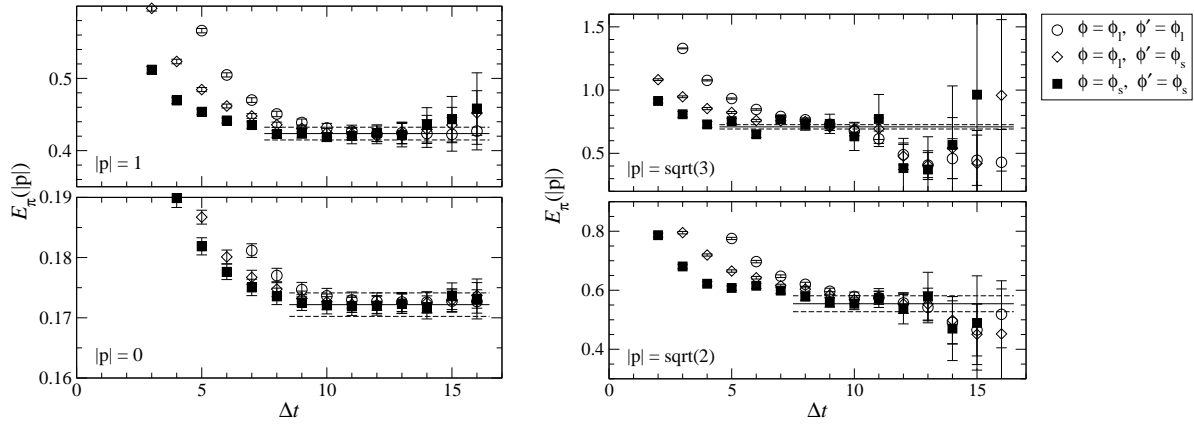

\begin{center}
\includegraphics[angle=0,width=0.41\linewidth,clip]%
                {em_msn_pp_mud0_mval00_tsrc00_set1.eps}
\hspace{3mm}
\includegraphics[angle=0,width=0.54\linewidth,clip]%
                {em_msn_pp_mud0_mval00_tsrc00_set2.eps}

\vspace{-3mm}
\caption{
   Effective value of pion energy $E_\pi(|\bfp|)$ at $(Q,m)\!=\!(0,0.015)$.
}
\label{fig:pff:Epi:m0015}
\end{center}
\end{figure}

The kinematical factor in Eq.~(\ref{eqn:pff:pff_v:dratio})
involves energies of the initial and final pion states.
Effective values of $E_\pi(|\bfp|)$ at our largest and smallest quark mass 
are plotted in Figs.~\ref{fig:pff:Epi:m0050} and \ref{fig:pff:Epi:m0015}.
Our data show clear signals up to the largest spatial momentum 
$|\bfp|\!\leq\!\sqrt{3}$,
since their statistical errors are greatly reduced 
by the use of the all-to-all propagators.
We determine $E_\pi(|\bfp|)$ 
by a single-cosh fit to $C_{\pi \pi, \phi_s \phi_s}(\Delta t;\bfp)$
with a fit range $\Delta t \in [\Delta t_{\rm min},N_t/2]$.
The minimum temporal separation $\Delta t_{\rm min}$ is chosen 
by inspecting $\Delta t_{\rm min}$ dependence of the fit result.
Numerical results for $E_\pi(|\bfp|)$ 
are summarized in Table~\ref{tbl:pff:Epi}.
They are consistent with the dispersion relation 
$E_\pi(|\bfp|)\!=\!\sqrt{M_\pi^2+\bfp^2}$,
which is commonly assumed in previous studies 
to estimate $E_\pi(|\bfp|)$ for $|\bfp|\!>\!0$.
In this work, however, 
we use the measured value of $E_\pi(|\bfp|)$
in order not to underestimate uncertainty in 
$E_\pi(|\bfp|)$ and hence $F_V(\Delta t,\Delta t^\prime;q^2)$.

\begin{ruledtabular}
\begin{table}[t]
\begin{center}
\caption{
   Fit results for pion energy $E_\pi(|\bfp|)$.
}
\label{tbl:pff:Epi}
\vspace{3mm}
\begin{tabular}{ll|lllllll}
   $Q$  & $m$    & $|\bfp|\!=\!0$        & $|\bfp|\!=\!1$ 
                 & $|\bfp|\!=\!\sqrt{2}$ & $|\bfp|\!=\!\sqrt{3}$ 
   \\ \hline
   0    & 0.015  & 0.1722(20)  & 0.4236(88)  & 0.554(27)   & 0.710(17)  \\
   0    & 0.025  & 0.2193(16)  & 0.4499(58)  & 0.5791(96)  & 0.694(16)  \\
   0    & 0.035  & 0.2610(15)  & 0.4706(33)  & 0.6091(59)  & 0.721(11)  \\
   0    & 0.050  & 0.3128(12)  & 0.5005(21)  & 0.6292(28)  & 0.732(11)  
   \\ \hline
   -2   & 0.050  & 0.3124(15)  & 0.4972(34)  & 0.6257(67)  & 0.738(14)  \\
   -4   & 0.050  & 0.3155(17)  & 0.4994(29)  & 0.6353(51)  & 0.705(10)  
\end{tabular}
\end{center}
\vspace{0mm}
\end{table}
\end{ruledtabular}

\begin{figure}[t]
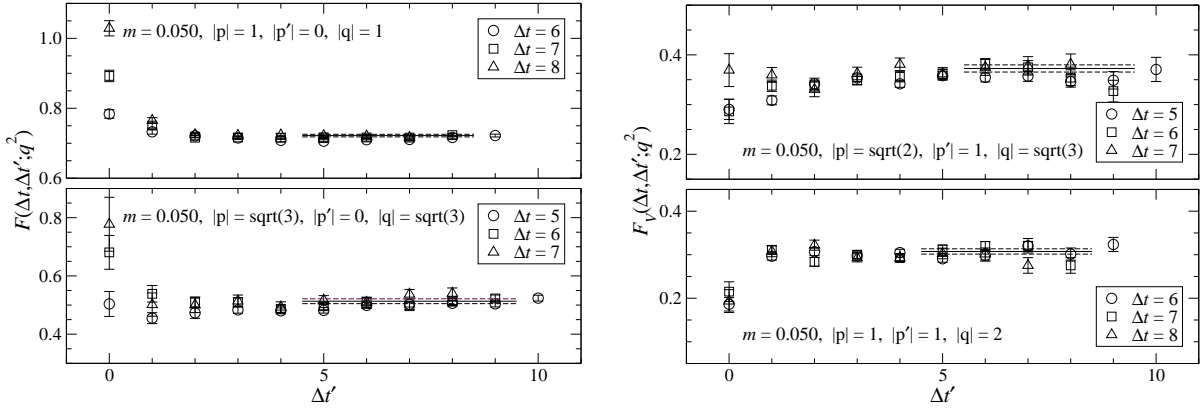

\begin{center}
\includegraphics[angle=0,width=0.48\linewidth,clip]%
                {pff_v4_m0050_mom0100_1900.eps}
\hspace{3mm}
\includegraphics[angle=0,width=0.48\linewidth,clip]%
                {pff_v4_m0050_mom0703_0101.eps}

\vspace{-3mm}
\caption{
   Effective value of vector form factor $F_V(\Delta t, \Delta t^\prime;q^2)$ 
   at $(Q,m)\!=\!(0,0.050)$.
   We only plot data with $\Delta t \! + \! \Delta t^\prime \! < \! N_t/2$.
   Momenta written in the legends are spatial and in units of $2\pi a/L$.
}
\label{fig:pff:pff_v:pff_eff:m0050}
\end{center}
\end{figure}

\begin{figure}[t]
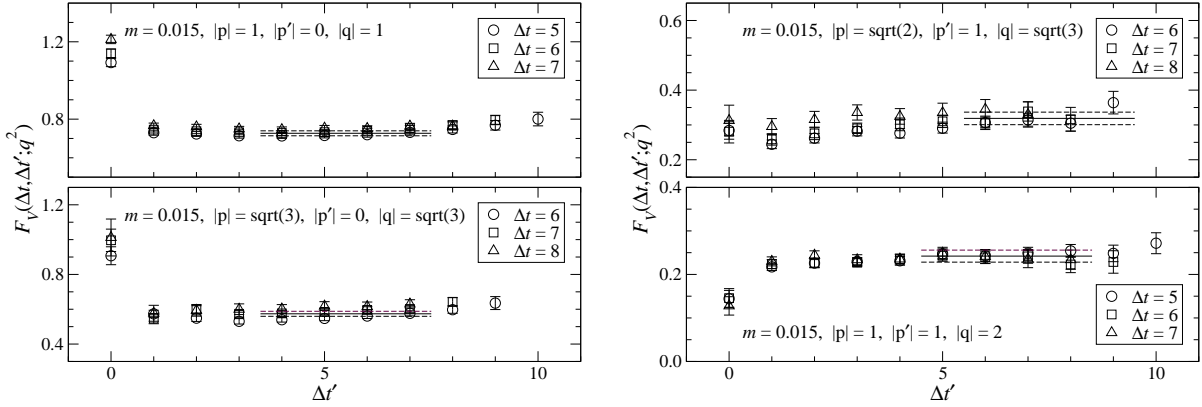

\begin{center}
\includegraphics[angle=0,width=0.48\linewidth,clip]%
                {pff_v4_m0015_mom0100_1900.eps}
\hspace{3mm}
\includegraphics[angle=0,width=0.48\linewidth,clip]%
                {pff_v4_m0015_mom0703_0101.eps}

\vspace{-3mm}
\caption{
   Effective value of $F_V(\Delta t, \Delta t^\prime;q^2)$ at $(Q,m)\!=\!(0,0.015)$.
}
\label{fig:pff:pff_v:pff_eff:m0015}
\end{center}
\end{figure}


We extract the vector form factor $F_V(q^2)$
by a constant fit to the effective value $F_V(\Delta t, \Delta t^\prime;q^2)$.
Examples of $F_V(\Delta t, \Delta t^\prime;q^2)$ 
are plotted in Figs.~\ref{fig:pff:pff_v:pff_eff:m0050} 
and \ref{fig:pff:pff_v:pff_eff:m0015}.
The all-to-all quark propagator enables us to 
change $\Delta t$ and $\Delta t^\prime$ independently 
in contrast to previous lattice studies with the point-to-all propagator
where their sum $\Delta t \! + \! \Delta t^\prime$ is kept fixed.
This is helpful to identify the plateau in $F_V(\Delta t, \Delta t^\prime;q^2)$
as well as to stabilize the fit by increasing the number of available data.
Fit results are summarized 
in Tables~\ref{tbl:pff:pff:Q0:m0050}\,--\,\ref{tbl:pff:pff:Q0:m0015}.
The statistical accuracy of $F_V(q^2)$ is typically 3\,--\,5\,\%
at all of our simulated quark masses.
Only the combination of our two largest momenta 
$(|\bfp|,|\bfp^\prime|)\!=\!(\sqrt{3},\sqrt{2})$
has larger statistical error of about 10\,\%.
This is probably because 
precise determination of the ground state contribution is difficult
due to very rapid damping of pion correlators with such large momenta. 

\begin{ruledtabular}
\begin{table}[t]
\begin{center}
\caption{
   Fit results for pion form factors at $(Q,m)\!=\!(0,0.050)$.
}
\label{tbl:pff:pff:Q0:m0050}
\begin{tabular}{lll|lll||lll|lll}
   $|\bfp|$  & $|\bfp^\prime|$  & $|\bfq|$  & $q^2$ 
                                            & $F_V(q^2)$ 
                                            & $\frac{F_S(q^2)}{F_S(q_{\rm ref}^2)}$ 
                                              \hspace{1mm}
                                            &
   $|\bfp|$  & $|\bfp^\prime|$  & $|\bfq|$  & $q^2$ 
                                            & $F_V(q^2)$ 
                                            & $\frac{F_S(q^2)}{F_S(q_{\rm ref}^2)}$ \\
   \hline
   $0$        & $0$             & $0$        & 0              & 1
                                                              & 1.30(11)   
                                                              &
   $\sqrt{3}$ & $0$             & $\sqrt{3}$ & $-$0.2865(95)  & 0.5132(83) 
                                                              & 0.705(44)
   \\
   $1$        & $0$             & $1$        & $-$0.1190(7)   & 0.7219(37) 
                                                              & --         
                                                              &
   $\sqrt{3}$ & $1$             & $\sqrt{2}$ & $-$0.2546(51)  & 0.522(17)  
                                                              & 0.728(66)
   \\
   $\sqrt{2}$ & $1$             & $1$        & $-$0.13763(68) & 0.660(13)  
                                                              & 0.918(37)  
                                                              & 
   $1$        & $1$             & $\sqrt{2}$ & $-$0.3084(0)   & 0.4755(47) 
                                                              & 0.759(21)
   \\
   $\sqrt{3}$ & $\sqrt{2}$      & $1$        & $-$0.1436(23)  & 0.599(49)  
                                                              & 0.907(90)  
                                                              &
   $\sqrt{2}$ & $1$             & $\sqrt{3}$ & $-$0.4461(7)   & 0.3726(72) 
                                                              & 0.681(34)
   \\
   $\sqrt{2}$ & $0$             & $\sqrt{2}$ & $-$0.2083(18)  & 0.5908(45) 
                                                              & 0.871(17)  
                                                              &
   $1$        & $1$             & $2$        & $-$0.6169(0)   & 0.3075(61) 
                                                              & 0.605(33)
\end{tabular}
\end{center}
\vspace{0mm}
\end{table}
\end{ruledtabular}
\begin{ruledtabular}
\begin{table}[t]
\begin{center}
\caption{
   Fit results for pion form factors at $(Q,m)\!=\!(0,0.035)$.
}
\label{tbl:pff:pff:Q0:m0035}
\begin{tabular}{lll|lll||lll|lll}
   $|\bfp|$  & $|\bfp^\prime|$  & $|\bfq|$  & $q^2$ 
                                            & $F_V(q^2)$ 
                                            & $\frac{F_S(q^2)}{F_S(q_{\rm ref}^2)}$ 
                                              \hspace{1mm}
                                            &
   $|\bfp|$  & $|\bfp^\prime|$  & $|\bfq|$  & $q^2$ 
                                            & $F_V(q^2)$ 
                                            & $\frac{F_S(q^2)}{F_S(q_{\rm ref}^2)}$ \\
   \hline
   $0$        & $0$             & $0$        & 0              & 1
                                                              & 1.28(10)
                                                              &
   $\sqrt{3}$ & $0$             & $\sqrt{3}$ & $-$0.2513(99)  & 0.5266(98)
                                                              & 0.673(69)
   \\
   $1$        & $0$             & $1$        & $-$0.1103(15)  & 0.7272(65)
                                                              & --
                                                              & 
   $\sqrt{3}$ & $1$             & $\sqrt{2}$ & $-$0.2459(54)  & 0.513(25)
                                                              & 0.783(66)
   \\
   $\sqrt{2}$ & $1$             & $1$        & $-$0.1350(17)  & 0.638(18)
                                                              & 0.902(43)
                                                              &
   $1$        & $1$             & $\sqrt{2}$ & $-$0.3084(0)   & 0.4518(80)
                                                              & 0.765(27)
   \\
   $\sqrt{3}$ & $\sqrt{2}$      & $1$        & $-$0.1417(27)  & 0.610(85)
                                                              & 0.890(87)
                                                              &
   $\sqrt{2}$ & $1$             & $\sqrt{3}$ & $-$0.4435(17)  & 0.358(11)
                                                              & 0.648(53)
   \\
   $\sqrt{2}$ & $0$             & $\sqrt{2}$ & $-$0.1873(44)  & 0.5964(75)
                                                              & 0.728(64)
                                                              &
   $1$        & $1$             & $2$        & $-$0.6169(0)   & 0.2882(76)
                                                              & 0.546(31)
\end{tabular}
\end{center}
\vspace{0mm}
\end{table}
\end{ruledtabular}
\begin{ruledtabular}
\begin{table}[t]
\begin{center}
\caption{
   Fit results for pion form factors at $(Q,m)\!=\!(0,0.025)$.
}
\label{tbl:pff:pff:Q0:m0025}
\begin{tabular}{lll|lll||lll|lll}
   $|\bfp|$  & $|\bfp^\prime|$  & $|\bfq|$  & $q^2$ 
                                            & $F_V(q^2)$ 
                                            & $\frac{F_S(q^2)}{F_S(q_{\rm ref}^2)}$ 
                                              \hspace{1mm}
                                            &
   $|\bfp|$  & $|\bfp^\prime|$  & $|\bfq|$  & $q^2$ 
                                            & $F_V(q^2)$ 
                                            & $\frac{F_S(q^2)}{F_S(q_{\rm ref}^2)}$ \\
   \hline
   $0$        & $0$             & $0$        & 0              & 1
                                                              & 1.29(13)
                                                              &
   $\sqrt{3}$ & $0$             & $\sqrt{3}$ & $-$0.237(15)   & 0.529(12)
                                                              & 0.650(35)
   \\
   $1$        & $0$             & $1$        & $-$0.1010(27)  & 0.7327(88)
                                                              & --
                                                              & 
   $\sqrt{3}$ & $1$             & $\sqrt{2}$ & $-$0.2489(83)  & 0.474(24)
                                                              & 0.56(11)
   \\
   $\sqrt{2}$ & $1$             & $1$        & $-$0.1375(27)  & 0.627(21)
                                                              & 0.939(61)
                                                              &
   $1$        & $1$             & $\sqrt{2}$ & $-$0.3084(0)   & 0.422(11)
                                                              & 0.686(42)
   \\
   $\sqrt{3}$ & $\sqrt{2}$      & $1$        & $-$0.1410(38)  & 0.550(60)
                                                              & 0.49(31)
                                                              &
   $\sqrt{2}$ & $1$             & $\sqrt{3}$ & $-$0.4460(27)  & 0.337(12)
                                                              & 0.548(50)
   \\
   $\sqrt{2}$ & $0$             & $\sqrt{2}$ & $-$0.1790(70)  & 0.600(10)
                                                              & 0.768(30)
                                                              &
   $1$        & $1$             & $2$        & $-$0.6169(0)   & 0.2561(87)
                                                              & 0.523(54)
\end{tabular}
\end{center}
\vspace{0mm}
\end{table}
\end{ruledtabular}
\begin{ruledtabular}
\begin{table}[t]
\begin{center}
\caption{
   Fit results for pion form factors at $(Q,m)\!=\!(0,0.015)$.
}
\label{tbl:pff:pff:Q0:m0015}
\begin{tabular}{lll|lll||lll|lll}
   $|\bfp|$  & $|\bfp^\prime|$  & $|\bfq|$  & $q^2$ 
                                            & $F_V(q^2)$ 
                                            & $\frac{F_S(q^2)}{F_S(q_{\rm ref}^2)}$ 
                                              \hspace{1mm}
                                            &
   $|\bfp|$  & $|\bfp^\prime|$  & $|\bfq|$  & $q^2$ 
                                            & $F_V(q^2)$ 
                                            & $\frac{F_S(q^2)}{F_S(q_{\rm ref}^2)}$ \\
   \hline
   $0$        & $0$             & $0$        & 0              & 1
                                                              & 1.29(19)
                                                              &
   $\sqrt{3}$ & $0$             & $\sqrt{3}$ & $-$0.174(18)   & 0.574(14)  
                                                              & 0.621(90)
   \\
   $1$        & $0$             & $1$        & $-$0.0910(44)  & 0.727(13)  
                                                              & --
                                                              &
   $\sqrt{3}$ & $1$             & $\sqrt{2}$ & $-$0.227(11)   & 0.484(33)  
                                                              & 0.80(12)
   \\
   $\sqrt{2}$ & $1$             & $1$        & $-$0.1372(73)  & 0.645(34)  
                                                              & 0.926(86)
                                                              &
   $1$        & $1$             & $\sqrt{2}$ & $-$0.3084(0)   & 0.403(16)  
                                                              & 0.643(95)
   \\ 
   $\sqrt{3}$ & $\sqrt{2}$      & $1$        & $-$0.1301(98)  & 0.629(74)  
                                                              & 0.81(41)
                                                              &
   $\sqrt{2}$ & $1$             & $\sqrt{3}$ & $-$0.4456(73)  & 0.319(18)  
                                                              & 0.599(81)
   \\
   $\sqrt{2}$ & $0$             & $\sqrt{2}$ & $-$0.162(21)   & 0.631(25)  
                                                              & 0.805(44)
                                                              &
   $1$        & $1$             & $2$        & $-$0.6169(0)   & 0.242(14)  
                                                              & 0.65(13)
\end{tabular}
\end{center}
\vspace{0mm}
\end{table}
\end{ruledtabular}


\begin{figure}[t]
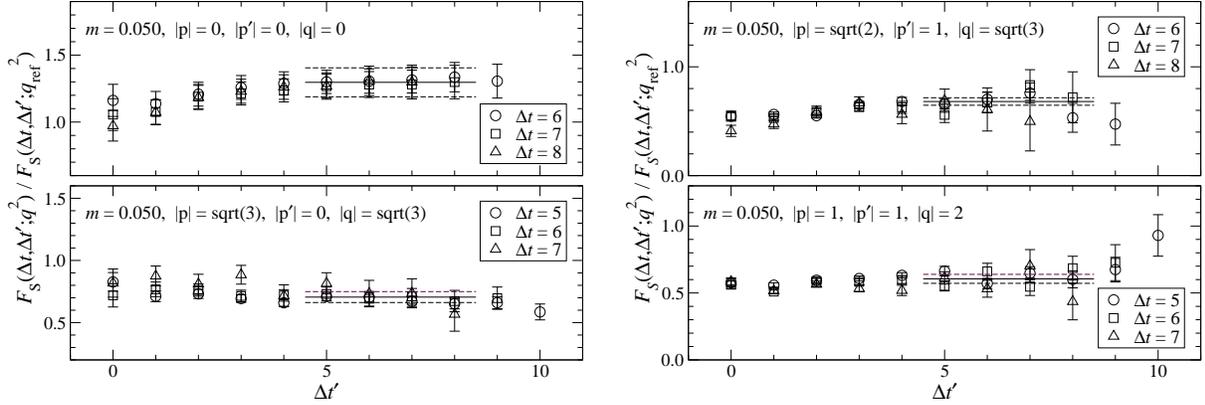

\begin{center}
\includegraphics[angle=0,width=0.48\linewidth,clip]%
                {pff_s_m0050_mom0000_1900.eps}
\hspace{3mm}
\includegraphics[angle=0,width=0.48\linewidth,clip]%
                {pff_s_m0050_mom0703_0101.eps}

\vspace{-3mm}
\caption{
   Effective value of normalized scalar form factor 
   $F_S(\Delta t, \Delta t^\prime;q^2)/F_S(\Delta t, \Delta t^\prime;q_{\rm ref}^2)$ at $(Q,m)\!=\!(0,0.050)$.
}
\label{fig:pff:pff_s:pff_eff:m0050}
\end{center}
\end{figure}

\begin{figure}[t]
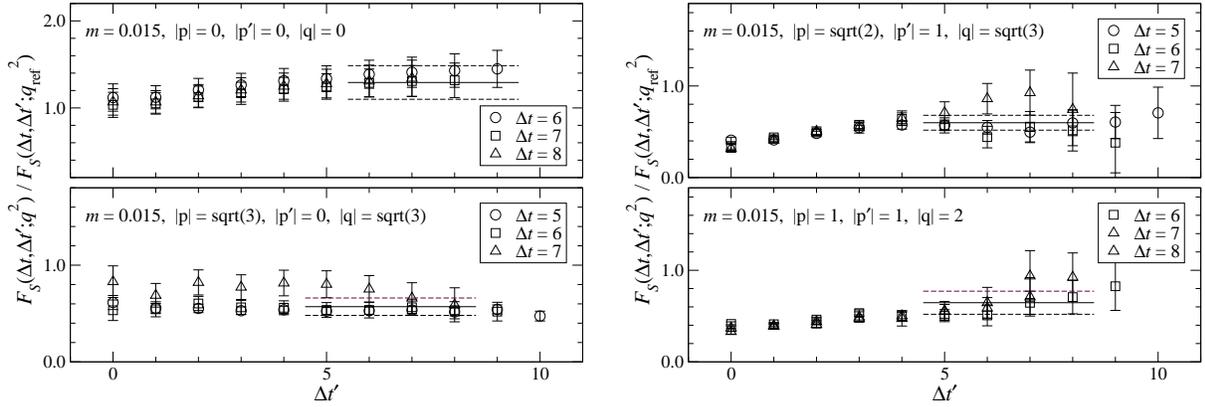

\begin{center}
\includegraphics[angle=0,width=0.48\linewidth,clip]%
                {pff_s_m0015_mom0000_1900.eps}
\hspace{3mm}
\includegraphics[angle=0,width=0.48\linewidth,clip]%
                {pff_s_m0015_mom0703_0101.eps}

\vspace{-3mm}
\caption{
   Effective value of 
   $F_S(\Delta t, \Delta t^\prime;q^2)/F_S(\Delta t, \Delta t^\prime;q_{\rm ref}^2)$ at $(Q,m)\!=\!(0,0.015)$.
}
\label{fig:pff:pff_s:pff_eff:m0015}
\end{center}
\end{figure}

\subsection{Scalar form factor}
\label{sec:pff:pff_s}

The scalar form factor normalized by the value at a certain momentum transfer 
$q_{\rm ref}^2$ can be calculated from the following ratio 
similar to that for $F_V(q^2)$
\bea
   \frac{F_S(\Delta t,\Delta t^\prime;q^2)}
        {F_S(\Delta t,\Delta t^\prime;q_{\rm ref}^2)}
   & = &
   \frac{R_S(\Delta t,\Delta t^\prime; q^2)}
        {R_S(\Delta t,\Delta t^\prime; q_{\rm ref}^2)},
   \label{eqn:pff:pff_s:dratio}
   \\ 
   R_S(\Delta t,\Delta t^\prime; q^2)
   & = & 
   \frac{1}{N_{|\bfp|,|\bfp^\prime|}} 
   \sum_{\mbox{\scriptsize fixed }|\bfp|,|\bfp^\prime|}
   \frac{C_{\pi 1 \pi}^{\rm (sngl)}(\Delta t,\Delta t^\prime; \bfp,\bfp^\prime)}
        {C_{\pi \pi, \phi_s \phi_l}(\Delta t;\bfp)\,
         C_{\pi \pi, \phi_l \phi_s}(\Delta t^\prime;\bfp^\prime)},
   \label{eqn:pff:pff_s:ratio}
\eea
where
\bea
   C_{\pi 1 \pi}^{\rm (sngl)}(\Delta t,\Delta t^\prime; \bfp,\bfp^\prime)
   & = & 
   C_{\pi 1 \pi}^{\rm (conn)}(\Delta t,\Delta t^\prime; \bfp,\bfp^\prime)
 - C_{\pi 1 \pi}^{\rm (disc)}(\Delta t,\Delta t^\prime; \bfp,\bfp^\prime)
   \\ \nn
   &&
   \hspace{36.5mm}
 + \, C_{\pi 1 \pi}^{\rm (vev)}(\Delta t,\Delta t^\prime; \bfp,\bfp^\prime)
   \label{eqn:pff:pff_s:sngl}
\eea
is the three-point function with the flavor-singlet scalar operator.
We note that $C_{\pi 1 \pi}^{\rm (disc)}\!-\!C_{\pi 1 \pi}^{\rm (vev)}$
suffers from a severe cancellation:
it is typically a subtraction of $O(1)$ quantities 
to extract their $O(10^{-3})$ difference.
Although this subtraction leads to a large statistical error in $F_S(q^2)$,
it is present only at $|\bfq|\!=\!0$,
since $C_{\pi 1 \pi}^{\rm (vev)}$ vanishes at nonzero $|\bfq|$.
In the following analysis, therefore, 
we use $F_S(q^2)$ normalized at the smallest nonzero $|q^2|$ 
with $|{\bf q}_{\rm ref}|\!=\!1$
rather than at $q^2\!=\!0$.
The effective value of the normalized scalar form factor
$F_S(\Delta t,\Delta t^\prime;q^2)/F_S(\Delta t,\Delta t^\prime;q_{\rm ref}^2)$
is plotted in Figs.~\ref{fig:pff:pff_s:pff_eff:m0050}
and \ref{fig:pff:pff_s:pff_eff:m0015}.
We summarize $F_S(q^2)/F_S(q_{\rm ref}^2)$ 
determined from a constant fit to 
$F_S(\Delta t,\Delta t^\prime;q^2)/F_S(\Delta t,\Delta t^\prime;q_{\rm ref}^2)$
in Tables~\ref{tbl:pff:pff:Q0:m0050}\,--\,\ref{tbl:pff:pff:Q0:m0015}.

\begin{figure}[t]
\begin{center}
\includegraphics[angle=0,width=0.48\linewidth,clip]%
                {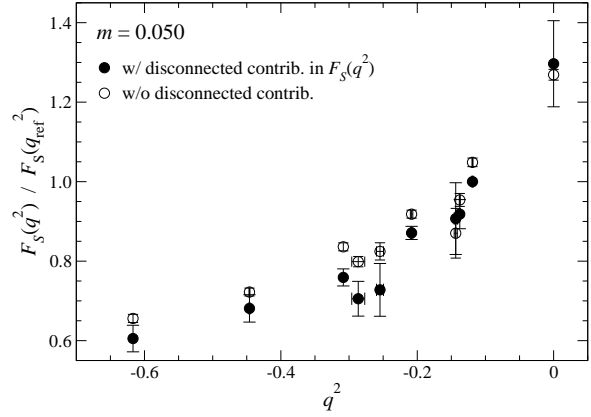}
\vspace{-3mm}
\caption{
   Normalized scalar form factor $F_S(q^2)/F_S(q_{\rm ref}^2)$ 
   with and without the contributions of disconnected diagram to $F_S(q^2)$.
   We use the normalization value $F_S(q_{\rm ref}^2)$ 
   with the disconnected contributions in both data.
}
\label{fig:pff:pff_s:disc}
\end{center}
\end{figure}

Figure~\ref{fig:pff:pff_s:disc}
compares $F_S(q^2)/F_S(q_{\rm ref}^2)$ to that without 
the contributions of the disconnected diagrams to $F_S(q^2)$.
We observe a small but significant deviation between the two data,
which implies the importance of the disconnected contributions.

An accurate estimate of $F_S(0)$ 
is useful for a precise determination of the scalar radius $\crad_S$,
as it characterizes the $q^2$ dependence of $F_S(q^2)$ near $q^2\!=\!0$.
Since the extraction of $F_S(0)$ from the three-point function suffers from 
large statistical error due to the subtraction of 
the vacuum expectation value (VEV) of the scalar operator,
we test an alternative calculation of $F_S(0)$ 
through the Feynman-Hellmann theorem
\bea
   F_S(0) 
   & = & 
   \frac{1}{2}\frac{\partial M_\pi^2}{\partial m},
   \label{eqn:pff:pff_s:fh}
\eea 
where the VEV subtraction is implicitly taken into account.
Note that the overall factor 1/2 is present in the RHS,
since $m$ is the mass of two degenerate quark flavors.
This and $F_S(q_{\rm ref})$ determined 
from a ratio of pion correlators 
\bea
   F_S(q^2) 
   & \xleftarrow[\Delta t,\Delta t^\prime\to \infty]{} & 
   2 Z_S E_\pi(|\bfp^\prime|)   
   \frac{C_{\pi 1 \pi}^{\rm (sngl)}(\Delta t,\Delta t^\prime; \bfp,\bfp^\prime)\,
         C_{\pi \pi, \phi_s \phi_l}(\Delta t^\prime; \bfp^\prime)}
        {C_{\pi \pi, \phi_s \phi_l}(\Delta t;\bfp)\,
         C_{\pi \pi, \phi_s \phi_s}(\Delta t+\Delta t^\prime;\bfp^\prime)}
   \label{eqn:pff:pff_s:drat2}
\eea
provide an alternative estimate of the normalized form factor 
$F_S(0)/F_S(q_{\rm ref}^2)$.
We determine $F_S(0)$ from the chiral fit of $M_\pi^2$ and 
$Z_S\!=\!0.838(14)(3)$ presented in Ref.~\cite{Spectrum:Nf2:RG+Ovr:JLQCD}.
Results for $F_S(0)$ and $F_S(0)/F_S(q_{\rm ref}^2)$
are summarized in Table~\ref{tbl:pff:pff_s:fh+drat2},
where the systematic error is estimated 
by changing the fitting form for $M_\pi^2$
and by taking account of the uncertainty in $Z_S$.
It turns out that, with the setup of our measurements,
the Feynman-Hellmann theorem (\ref{eqn:pff:pff_s:fh})
leads to a slightly smaller uncertainty in $F_S(0)/F_S(q_{\rm ref}^2)$
than that from the ratio (\ref{eqn:pff:pff_s:dratio}).
%
We therefore 
adopt $F_S(0)/F_S(q_{\rm ref}^2)$ in Table~\ref{tbl:pff:pff_s:fh+drat2}
and $F_S(q^2\! \ne \! 0)/F_S(q_{\rm ref}^2)$ in 
Tables~\ref{tbl:pff:pff:Q0:m0050}\,--\,\ref{tbl:pff:pff:Q0:m0015}
in the following analysis.

\begin{ruledtabular}
\begin{table}[t]
\begin{center}
\caption{
   Scalar form factor $F_S(0)$ 
   determined from Feynman-Hellmann theorem Eq.~(\ref{eqn:pff:pff_s:fh}),
   and its value normalized 
   by $F_S(q_{\rm ref}^2)$ from ratio (\ref{eqn:pff:pff_s:drat2}).
   The first error is statistical. The second is systematics 
   due to the choice of the fit form for $M_\pi^2$ and uncertainty in $Z_S$.
}
\label{tbl:pff:pff_s:fh+drat2}
\begin{tabular}{l|llll}
   $m$\hspace{22mm} & 0.015\hspace{15mm} & 0.025\hspace{15mm}
                    & 0.035\hspace{15mm} & 0.050\hspace{15mm}
   \\ \hline
   $F_S(0)$ &
   1.149(6)(22)  & 1.162(18)(37) & 1.208(29)(47) & 1.319(44)(52)
   \\
   $F_S(0)/F_S(q_{\rm ref}^2)$ &
   1.413(70)(28) & 1.415(44)(44) & 1.338(38)(51) & 1.441(53)(58) 
\end{tabular}
\end{center}
\vspace{0mm}
\end{table}
\end{ruledtabular}

\subsection{Finite volume correction}

The finite volume effect could be significant 
at two smallest quark masses $m\!=\!0.015$ and 0.025,
as the value of $M_\pi\,L$ is less than 4.
We estimate the finite volume correction (FVC) to the pion mass 
as presented in Ref.~\cite{Spectrum:Nf2:RG+Ovr:JLQCD}. 
The FVC to the vector form factor $F_V(q^2)$
has been calculated within one-loop ChPT 
in Refs.~\cite{FVC:PFF:V:ChPT,FVC:PFF:V:LChPT}
by replacing the loop integral by a discrete sum
\bea
   \Delta F_V(q^2)
   & = &
   \frac{1}{2F^2}
   \int_0^1 dx 
      \left\{ 
         I_{1/2}(L,x\bfq,M_\pi^2+x(1-x)q^2) - I_{1/2}(L,{\bf 0}, M_\pi^2)
      \right\},
   \label{pff:fvc:pff_v}
\eea
where
\bea
   I_s(L,\Delta \bfk, M^2)
   & = &
   \int \frac{d^3 k}{(2\pi)^3} 
      \frac{1}{\left\{ (\bfk + \Delta \bfk)^2 + M^2 \right\}^s}
   -
   \frac{1}{L^3} \sum_{\bfk} 
      \frac{1}{\left\{ (\bfk + \Delta \bfk)^2 + M^2 \right\}^s}.
   \label{eqn:pff:fvc:Is}
\eea
This function can be evaluated through the elliptic $\theta$ function
\cite{FVC:PFF:V:Is:1,FVC:PFF:V:Is:2} as 
\bea
   I_s(L,\Delta \bfk, M^2)
   & = & 
   \frac{1}{(4\pi)^{3/2}\Gamma(s)} 
   \int_0^\infty d\tau 
      \tau ^{s-5/2} e^{-M^2\tau}
      \left\{ 
         1 -\prod_{i=1}^3 \theta_3 \left(
                                      \frac{L}{2}\Delta k_i, e^{-L^2/4\tau} 
                                   \right)
      \right\},
   \label{eqn:pff:fvc:Is:theta3}
   \\
   \theta_3(u,q)
   & = &
   \sum_{n=-\infty}^{\infty} q^{n^2} e^{2nui}.
   \label{eqn:pff:fvc:theta3}
\eea  
The FVC to the scalar form factor $F_S(q^2)$ is similarly evaluated as 
\bea
   \Delta F_S(q^2)
   & = &
   \frac{B}{2F^2}
   \int_0^1 dx 
      \left\{ 
         - \frac{2q^2+M_\pi^2}{2} I_{3/2}(L,x\bfq,M_\pi^2+x(1-x)q^2) 
         + I_{1/2}(L,{\bf 0}, M_\pi^2)
      \right\},
   \label{pff:fvc:pff_s}
\eea
where we use $B\!=\!2.10(45)$~GeV 
obtained from our analysis of the pion mass \cite{Spectrum:Nf2:RG+Ovr:JLQCD}.
We find that $\Delta F_V(q^2)$ has a mild $q^2$ dependence and 
its magnitude is similar to or smaller than the statistical error:
it is typically 3\,--\,5\,\% at $m\!=\!0.015$, and 
decreases down to $\lesssim 1$\,\% at $m\!=\!0.050$.
The FVC to the normalized scalar form factor $F_S(q^2)/F_S(q_{\rm ref}^2)$ 
is below the statistical uncertainty even at our smallest quark mass
due to a partial cancellation of FVCs in the ratio.
We use $F_V(q^2)$ and $F_S(q^2)/F_S(q_{\rm ref}^2)$ 
with the FVC included in the following analysis.

\begin{figure}[t]
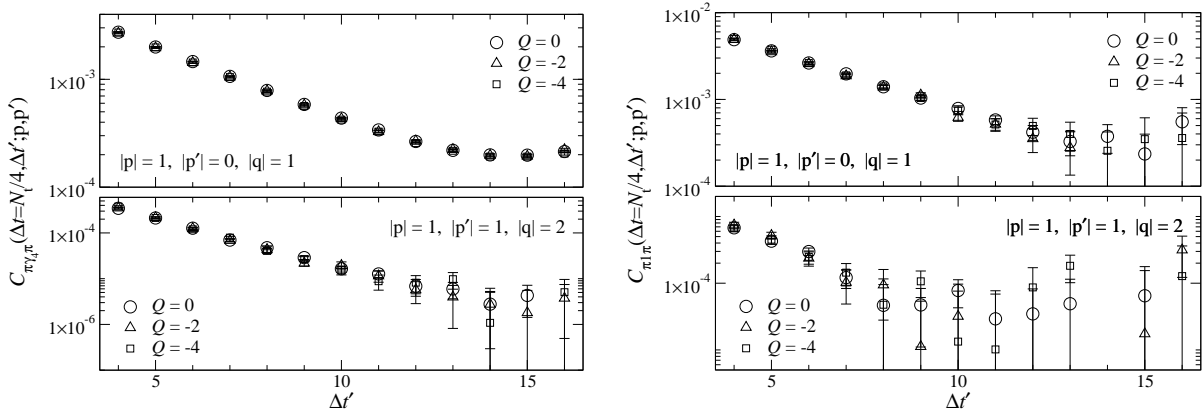

\begin{center}
\includegraphics[angle=0,width=0.48\linewidth,clip]%
                {msn_3pt_p-p-v4_mud3_mval333_mom0100_smr11.eps}
\hspace{3mm}
\includegraphics[angle=0,width=0.48\linewidth,clip]%
                {msn_3pt_sngl_p-p-s_mud3_mval333_mom0100_smr11.eps}

\vspace{-3mm}
\caption{
   Three-point functions with vector current (left panels) and 
   scalar operator (right panels) 
   calculated in different topological sectors for $m\!=\!0.050$.
   Top and bottom panels show data at the smallest and largest
   nonzero $|q^2|$.
}
\label{fig:pff:fixedQ:msn_corr_3pt}
\end{center}
\end{figure}

\subsection{Fixed topology effects}
\label{sec:pff:fixedQ}

The form factors listed 
in Tables~\ref{tbl:pff:pff:Q0:m0050}\,--\,\ref{tbl:pff:pff:Q0:m0015}
are extracted in the trivial topological sector,
and are subject to effects of the fixed topology.
The effects are known to be suppressed by the inverse space-time volume $1/V$ 
\cite{fixed_Q:BCNW,fixed_Q:AFHO}, 
and are systematically correctable \cite{fixed_Q:AFHO}. 
%
In order to confirm 
if the correction to our data is statistically insignificant,
we repeat the calculation of $F_V(q^2)$ and $F_S(q^2)$ 
in non-trivial topological sectors with $Q\!=\!-2$ and $-4$ at $m\!=\!0.050$.
Note that, LO ChPT predicts 
$\langle Q^2 \rangle \! = \! m \Sigma V / 2 \! \sim \!  8$
with our estimate of the chiral condensate 
$\Sigma^{\scriptsize \overline{\mbox{MS}}}(\mbox{2~GeV})\!=\!(0.236(^{+14}_{-5})~\mbox{GeV})^3$ \cite{Spectrum:Nf2:RG+Ovr:JLQCD}.
It is therefore not necessary to simulate topological sectors 
with $|Q| \gg \sqrt{\langle Q^2 \rangle} \sim 3$ at this quark mass.

\begin{ruledtabular}
\begin{table}[t]
\begin{center}
\caption{
   Fit results for pion form factors at $(Q,m)\!=\!(-2,0.050)$.
}
\label{tbl:pff:pff:Q-2:m0050}
\begin{tabular}{lll|lll||lll|lll}
   $|\bfp|$  & $|\bfp^\prime|$  & $|\bfq|$  & $q^2$ 
                                            & $F_V(q^2)$ 
                                            & $\frac{F_S(q^2)}{F_S(q_{\rm ref}^2)}$ 
                                              \hspace{1mm}
                                            &
   $|\bfp|$  & $|\bfp^\prime|$  & $|\bfq|$  & $q^2$ 
                                            & $F_V(q^2)$ 
                                            & $\frac{F_S(q^2)}{F_S(q_{\rm ref}^2)}$ \\
   \hline
   $1$        & $0$             & $1$        & $-$0.1200(11)  & 0.7145(46)
                                                              & --         
                                                              &
   $\sqrt{3}$ & $1$             & $\sqrt{2}$ & $-$0.2503(70)  & 0.481(22)
                                                              & 0.79(13)
   \\
   $\sqrt{2}$ & $1$             & $1$        & $-$0.1377(15)  & 0.667(20)
                                                              & 0.923(51)
                                                              & 
   $1$        & $1$             & $\sqrt{2}$ & $-$0.3084(0)   & 0.4640(65)
                                                              & 0.738(29)
   \\
   $\sqrt{3}$ & $\sqrt{2}$      & $1$        & $-$0.1416(33)  & 0.662(56)
                                                              & 0.32(34)
                                                              &
   $\sqrt{2}$ & $1$             & $\sqrt{3}$ & $-$0.4461(15)  & 0.365(11)
                                                              & 0.694(46)
   \\
   $\sqrt{2}$ & $0$             & $\sqrt{2}$ & $-$0.2102(40)  & 0.5910(75)
                                                              & 0.829(23)
                                                              &
   $1$        & $1$             & $2$        & $-$0.6169(0)   & 0.2962(89)
                                                              & 0.635(55)
   \\
   $\sqrt{3}$ & $0$             & $\sqrt{3}$ & $-$0.281(12)   & 0.5113(97)
                                                              & 0.785(65)
                                                              &
   --         & --              & --         & --             & --
                                                              & --
\end{tabular}
\end{center}
\end{table}
\end{ruledtabular}

\begin{ruledtabular}
\begin{table}[t]
\begin{center}
\caption{
   Fit results for pion form factors at $(Q,m)\!=\!(-4,0.050)$.
}
\label{tbl:pff:pff:Q-4:m0050}
\begin{tabular}{lll|lll||lll|lll}
   $|\bfp|$  & $|\bfp^\prime|$  & $|\bfq|$  & $q^2$ 
                                            & $F_V(q^2)$ 
                                            & $\frac{F_S(q^2)}{F_S(q_{\rm ref}^2)}$ 
                                              \hspace{1mm}
                                            &
   $|\bfp|$  & $|\bfp^\prime|$  & $|\bfq|$  & $q^2$ 
                                            & $F_V(q^2)$ 
                                            & $\frac{F_S(q^2)}{F_S(q_{\rm ref}^2)}$ \\
   \hline
   $1$        & $0$             & $1$        & $-$0.1204(9)   & 0.7234(58)
                                                              & --         
                                                              &
   $\sqrt{3}$ & $1$             & $\sqrt{2}$ & $-$0.2660(42)  & 0.561(22)
                                                              & 0.94(11)
   \\
   $\sqrt{2}$ & $1$             & $1$        & $-$0.1357(11)  & 0.641(21)
                                                              & 0.899(49)
                                                              & 
   $1$        & $1$             & $\sqrt{2}$ & $-$0.3084(0)   & 0.4810(83)
                                                              & 0.846(29)
   \\
   $\sqrt{3}$ & $\sqrt{2}$      & $1$        & $-$0.1493(16)  & 0.699(76)
                                                              & 0.39(32)
                                                              &
   $\sqrt{2}$ & $1$             & $\sqrt{3}$ & $-$0.4442(11)  & 0.3714(96)
                                                              & 0.754(57)
   \\
   $\sqrt{2}$ & $0$             & $\sqrt{2}$ & $-$0.2062(29)  & 0.5875(48)
                                                              & 0.872(26)
                                                              &
   $1$        & $1$             & $2$        & $-$0.6169(0)   & 0.3143(86)
                                                              & 0.603(42)
   \\
   $\sqrt{3}$ & $0$             & $\sqrt{3}$ & $-$0.3107(78)  & 0.5392(87)
                                                              & 0.820(52)
                                                              &
   --         & --              & --         & --             & --
                                                              & --
\end{tabular}
\end{center}
\vspace{0mm}
\end{table}
\end{ruledtabular}

\begin{figure}[t]
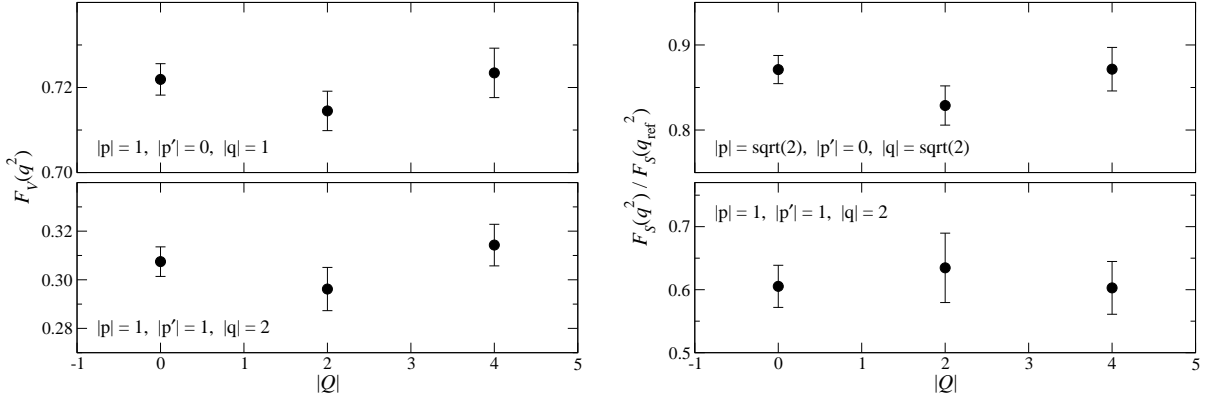

\begin{center}
\includegraphics[angle=0,width=0.48\linewidth,clip]%
                {pff_v4_vs_Q.eps}
\hspace{3mm}
\includegraphics[angle=0,width=0.48\linewidth,clip]%
                {pff_s_vs_Q.eps}

\vspace{-3mm}
\caption{
   Vector (left panels) and scalar form factors (right panels) 
   calculated in different topological sectors for $m\!=\!0.050$.
}
\label{fig:pff:fixedQ:pff_v_s}
\end{center}
\end{figure}

We compare three-point functions calculated in the different topological 
sectors in Fig.~\ref{fig:pff:fixedQ:msn_corr_3pt}, 
where no systematic deviation among the data is observed.
This is also the case for $F_{V,S}(q^2)$ 
summarized in Tables~\ref{tbl:pff:pff:Q-2:m0050}
and \ref{tbl:pff:pff:Q-4:m0050}.
The form factors at $Q\!=\!0$, $-$2 and $-$4 are consistent 
with each other within two standard deviations 
as shown in Fig.~\ref{fig:pff:fixedQ:pff_v_s}.
Although the effect of the fixed topology is likely 
below our statistical accuracy, 
we take the spread in $F_{V,S}(q^2)$ as a systematic error at $m\!=\!0.050$
and an uncertainty of the same magnitude is assumed at $m\!<\!0.050$
in the following analysis.

%% file: 4.q2_interp.tex
\section{Parametrization of the $q^2$ dependence}
\label{sec:q2_interp}


\subsection{Vector form factor}

\begin{figure}[t]
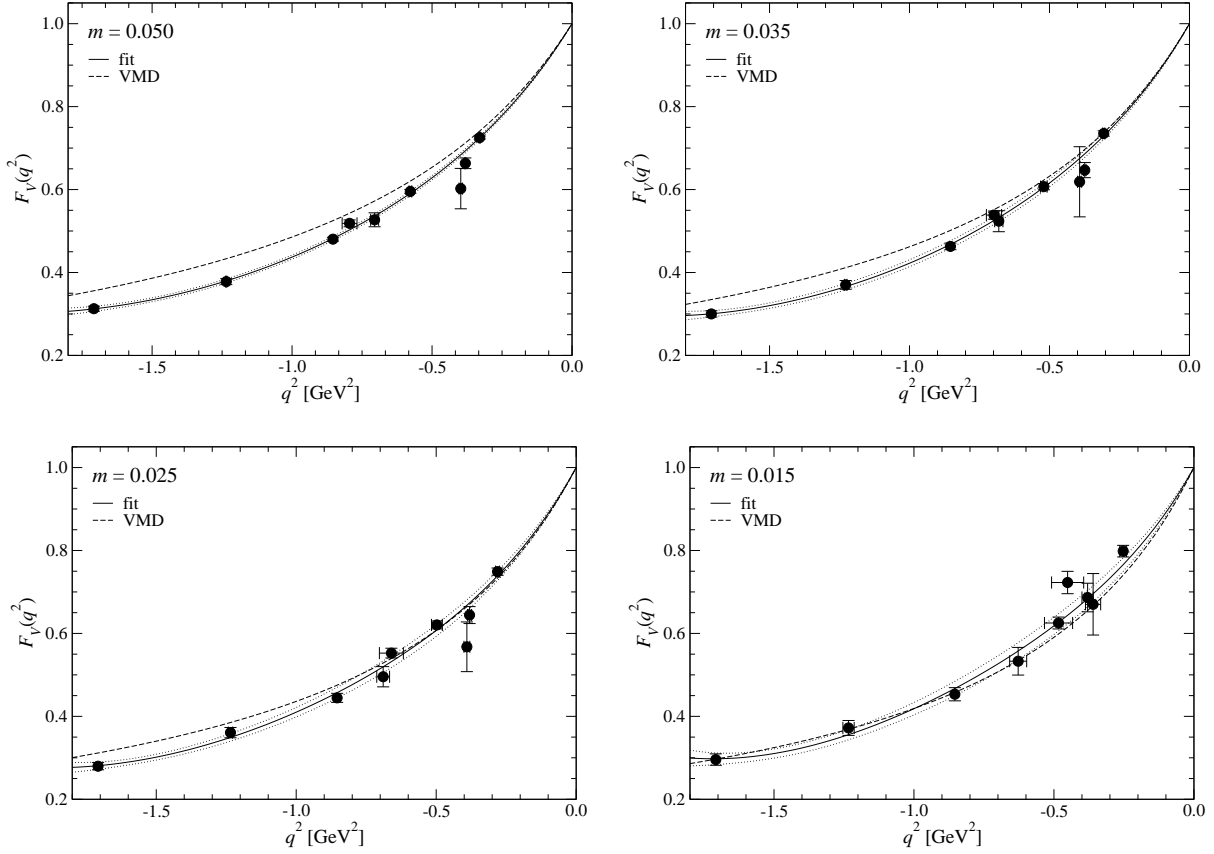

\begin{center}
\includegraphics[angle=0,width=0.48\linewidth,clip]{pff_v_vs_q2_m0050.phys.eps}
\hspace{3mm}
\includegraphics[angle=0,width=0.48\linewidth,clip]{pff_v_vs_q2_m0035.phys.eps}
\vspace{5mm}

\includegraphics[angle=0,width=0.48\linewidth,clip]{pff_v_vs_q2_m0025.phys.eps}
\hspace{3mm}
\includegraphics[angle=0,width=0.48\linewidth,clip]{pff_v_vs_q2_m0015.phys.eps}

\vspace{-3mm}
\caption{
   Vector form factor $F_V(q^2)$ as a function of $q^2$. 
   Solid and dotted lines show the fit curve 
   (\ref{eqn:q2_interp:vs_q2:pff_v}) and its error. 
   The $q^2$ dependence from the vector meson dominance model 
   is shown by the dashed line.
}
\label{fig:q2_interp:pff_v}
\end{center}
\end{figure}

The vector form factor $F_V(q^2)$ is plotted as a function of $q^2$ 
in Fig.~\ref{fig:q2_interp:pff_v}.
Its $q^2$ dependence turns out to be close to 
the expectation from the vector meson dominance (VMD) hypothesis
\bea
   F_V(q^2) 
   & = &
   \frac{1}{1-q^2/M_\rho^2},
   \label{eqn:q2_interp:vmd}
\eea
where $M_\rho$ is the vector meson mass calculated at the same quark mass.
%
%
Then, we assume that
the small deviation due to higher poles or cuts
can be well parametrized by a polynomial form.
We therefore fit our data to the following form 
\bea
   F_V(q^2)
   & = & 
   \frac{1}{1-q^2/M_{\rho}^2} + a_{V,1}\,q^2 + a_{V,2}\,(q^2)^2 
                              + a_{V,3}\,(q^2)^3
   \label{eqn:q2_interp:vs_q2:pff_v}
\eea
in order to extract the charge radius $\crad_V$ and the curvature $c_V$
\bea
   \crad_V  = \left. 
                 6\frac{\partial F_V(q^2)}{\partial (q^2)} 
              \right|_{q^2=0},
   \hspace{10mm}
   c_V      = \left. 
                 \frac{\partial^2 F_V(q^2)}{\partial (q^2)^2} 
              \right|_{q^2=0}.
\eea
The fit curve is plotted in Fig.~\ref{fig:q2_interp:pff_v} 
and numerical results are summarized in Table~\ref{tbl:q2_interp:vs_q2:pff_v}.
The fit describes our data reasonably well, 
and results for $\crad_V$ and $c_V$ do not change significantly 
by including or excluding the cubic term $a_{V,3}(q^2)^3$.
In the following analysis,
we employ results from the parametrization with the cubic term.


\begin{ruledtabular}
\begin{table}[t]
\begin{center}
\caption{
   Parametrization Eq.~(\ref{eqn:q2_interp:vs_q2:pff_v}) 
   for vector form factor $F_V(q^2)$. 
   Results for the vector radius $\crad_V$ and curvature $c_V$ in lattice units
   are also listed.
}
\label{tbl:q2_interp:vs_q2:pff_v}
\begin{tabular}{l|llll||ll}
   $m$   & $\chi^2/{\rm d.o.f.}$ & $a_{V,1}$ & $a_{V,2}$ & $a_{V,3}$ 
                                 & $\crad_V$ & $c_V$ \\
   \hline
   0.050 & 1.8(0.8) &  0.187(33) &  0.181(55)   & --        
                    & 18.72(20)  &  8.786(55)   \\
   0.050 & 1.7(0.9) &  0.116(64) &  $-$0.22(26) & $-$0.48(28) 
                    & 18.30(38)  &  8.38(26)  
   \\ \hline
   0.035 & 1.3(0.5) &  0.136(60) &  0.124(96)   & --
                    & 20.15(36)  & 10.51(10)   \\
   0.035 & 1.2(0.7) &  0.02(11)  &  $-$0.50(44) & $-$0.73(46)
                    & 19.48(66)  &  9.88(44)
   \\ \hline
   0.025 & 1.9(0.6) &  0.034(90)   & $-$0.03(14)  & -- 
                    & 21.69(54)    & 12.78(14)    \\
   0.025 & 1.7(0.7) & $-$0.14(16)  & $-$0.98(59)  & $-$1.10(56) 
                    & 20.64(98)    & 11.83(59) 
   \\ \hline
   0.015 & 0.8(0.8) & $-$0.09(12)  & $-$0.15(19)  & --
                    & 22.51(72)    & 14.58(19)    \\
   0.015 & 0.5(0.8) & $-$0.35(25)  & $-$1.56(96)  & $-$1.60(95) 
                    & 20.9(1.5)    & 13.18(96) 
\end{tabular}
\end{center}
\vspace{5mm}
\end{table}
\end{ruledtabular}

\begin{figure}[t]
\begin{center}
\includegraphics[angle=0,width=0.50\linewidth,clip]{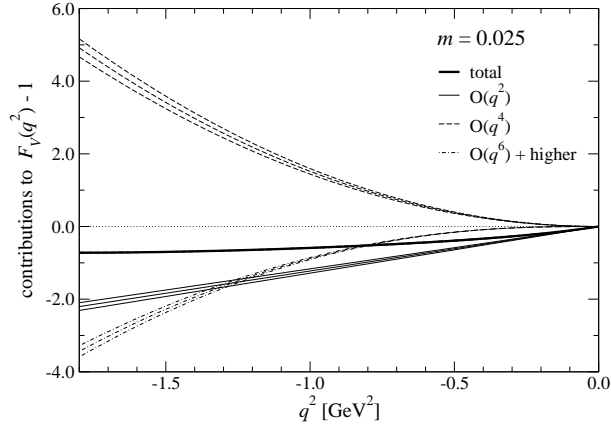}

\vspace{-3mm}
\caption{
   Contributions in the $q^2$ expansion of $F_V(q^2)\!-\!1$ at $m\!=\!0.025$. 
   Thin solid, dashed and dot-dashed lines show $O(q^2)$, $O(q^4)$ and 
   higher contributions, 
   whereas the thick solid line is their total.
}
\label{fig:q2_interp:pff_v:contrib}
\end{center}
\end{figure}

One of the main purposes of this work is 
to investigate whether the $q^2$ dependence of our data 
can be described by two-loop ChPT \cite{PFF_V:ChPT:NNLO:2,PFF_V:ChPT:NNLO}.
Figure~\ref{fig:q2_interp:pff_v:contrib} shows 
contributions to $F_V(q^2)$ from each order $(q^2)^n$ of 
a Taylor expansion of Eq.~(\ref{eqn:q2_interp:vs_q2:pff_v}).
We find that
$O(q^6)$ and higher order contributions to $F_V(q^2)$ are sufficiently 
small only below $|q^2|\!\simeq\!0.3~\mbox{GeV}^2$,
which is however around our smallest value of $|q^2|$.
We therefore do not use the parametrization based on ChPT in this study.
Note that such large higher order contributions are unavoidable 
unless $|q^2| \! \ll \! M_\rho^2$,
because the VMD form is a good approximation of $F_V(q^2)$.

\begin{ruledtabular}
\begin{table}[t]
\begin{center}
\caption{
   Single pole fit (\ref{eqn:q2_interp:vs_q2:pff_v:pole}). 
   We also list the $\rho$ meson mass at simulated quark masses.
}
\label{tbl:q2_interp:vs_q2:pff_v:pole}
\begin{tabular}{l|llll||ll}
   $m$   & $\chi^2/{\rm d.o.f.}$ & $M_{\rm pole}$ & $\crad_V$ & $c_V$ & $M_\rho$
   \\ \hline
   0.050 & 4.3(1.1) & 0.5431(50) & 20.34(38) & 11.49(43) & 0.5839(28) 
   \\ \hline
   0.035 & 2.3(0.7) & 0.5270(80) & 21.61(65) & 12.97(78) & 0.5570(31) 
   \\ \hline
   0.025 & 2.9(0.8) & 0.511(11)  & 23.0(1.0) & 14.6(1.3) & 0.5285(43) 
   \\ \hline
   0.015 & 0.7(0.7) & 0.520(16)  & 22.2(1.4) & 13.7(1.7) & 0.5104(55)  
\end{tabular}
\end{center}
\vspace{0mm}
\end{table}
\end{ruledtabular}

We also test a single pole ansatz often used in the previous studies 
\bea
   F_V(q^2)
   & = & 
   \frac{1}{1-q^2/M_{\rm pole}^2}.
   \label{eqn:q2_interp:vs_q2:pff_v:pole}
\eea
As summarized in Table~\ref{tbl:q2_interp:vs_q2:pff_v:pole},
this fit tends to give a slightly higher $\chi^2$ and $\crad_V$
than those from Eq.~(\ref{eqn:q2_interp:vs_q2:pff_v}).
This may suggest that it is difficult to describe our precise data of 
$F_V(q^2)$ in the whole region of $q^2[\mbox{GeV}^2] \! \in \! [-1.7,0]$ 
by a simple pole-dominance form.


\subsection{Scalar form factor}
\label{sec:q2_interp:pff_s}

\begin{figure}[t]
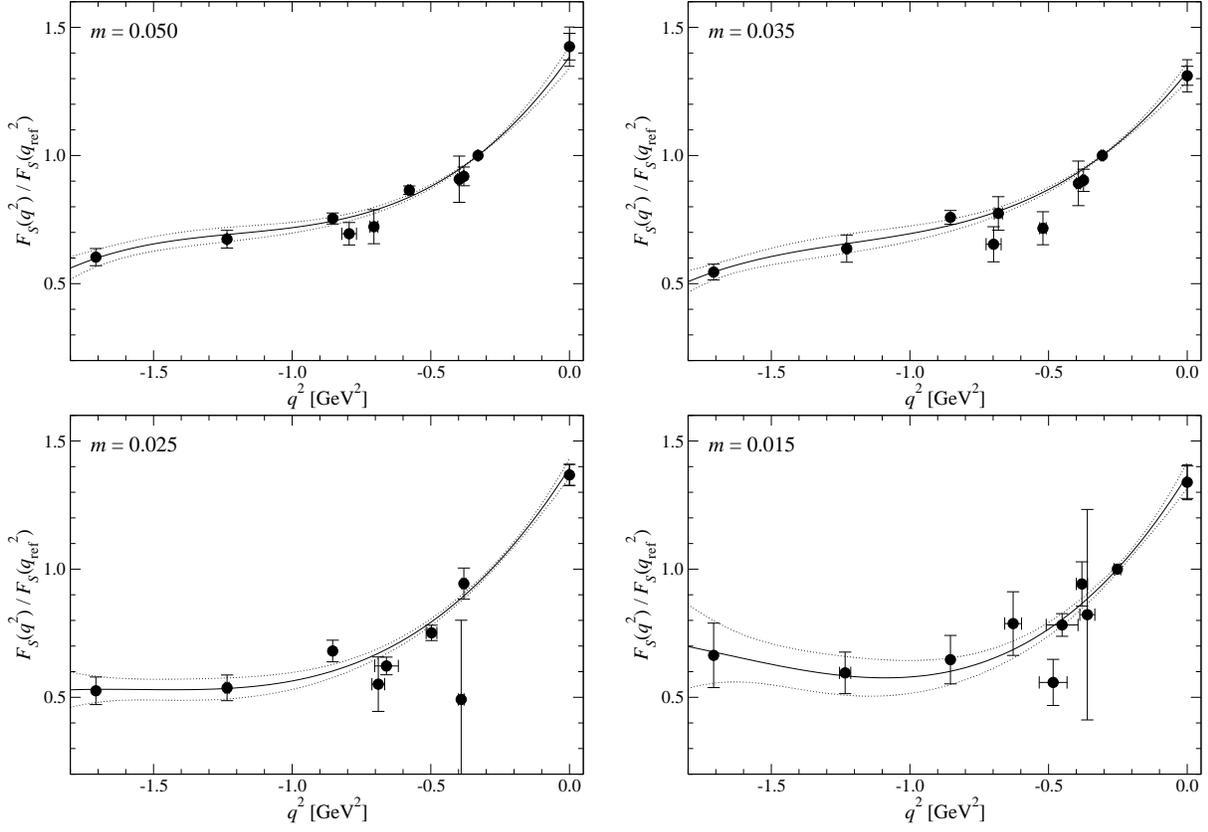

\begin{center}
\includegraphics[angle=0,width=0.48\linewidth,clip]{pff_s_vs_q2_norm-mom0100_m0050.phys.eps}
\hspace{3mm}
\includegraphics[angle=0,width=0.48\linewidth,clip]{pff_s_vs_q2_norm-mom0100_m0035.phys.eps}

\includegraphics[angle=0,width=0.48\linewidth,clip]{pff_s_vs_q2_norm-mom0100_m0025.phys.eps}
\hspace{3mm}
\includegraphics[angle=0,width=0.48\linewidth,clip]{pff_s_vs_q2_norm-mom0100_m0015.phys.eps}

\vspace{-3mm}
\caption{
   Normalized scalar form factor $F_S(q^2)/F_S(q_{\rm ref}^2)$ 
   as a function of $q^2$. 
   Solid and dotted lines show the cubic fit and its error. 
}
\label{fig:q2_interp:pff_s}
\end{center}
\end{figure}

\begin{ruledtabular}
\begin{table}[t]
\begin{center}
\caption{
   Parametrization Eq.~(\ref{eqn:q2_interp:vs_q2:pff_s}) 
   for scalar form factor $F_S(q^2)$. 
   Results for the vector radius $\crad_S$ and curvature $c_S$ in lattice units
   are also listed.
}
\label{tbl:q2_interp:vs_q2:pff_s}
\begin{tabular}{l|lllll||ll}
   $m$   & $\chi^2/{\rm d.o.f.}$ & $a_{S,1}$ & $a_{S,2}$ 
                                 & $a_{S,3}$ & $a_{S,4}$ 
                                 & $\crad_S$ & $c_S$   \\
   \hline
   0.050 & 1.3(1.0) & 3.04(31)   &  6.6(1.3)  &  5.1(1.4) & --
                    & 18.2(1.9)  &  6.6(1.3)  \\
   0.050 & 1.3(1.1) & 3.51(49)   & 10.5(3.5)  &  1.6(9.2) & 9.6(7.7)  
                    & 21.1(3.0)  & 10.5(3.5)  
   \\ \hline
   0.035 & 1.8(1.0) & 2.79(31)   &  5.6(1.4)  &  4.3(1.6) & -- 
                    & 16.8(1.9)  &  5.6(1.4)  \\
   0.035 & 1.9(1.2) & 2.60(51)   &  3.5(4.8)  & $-$3(15)  &  $-$6(13)
                    & 15.6(3.1)  &  3.5(4.8)  
   \\ \hline
   0.025 & 1.9(1.1) & 3.37(32)   &  6.1(1.6)  &    3.6(1.8) & --
                    & 20.2(1.9)  &  6.1(1.6)  \\
   0.025 & 1.9(1.3) & 2.97(53)   &  1.8(4.6)  & $-$9(13)    & $-$12(11) 
                    & 17.8(3.2)  &  1.8(4.6)  
   \\ \hline
   0.015 & 1.5(1.0) & 3.51(51)   &  6.6(2.7)  &  3.6(3.3)   & --
                    & 21.0(3.1)  &  6.6(2.7)  \\
   0.015 & 1.7(1.1) & 3.0(1.0)   &  1.0(9.1)  & $-$14(26)   & $-$16(22)
                    & 18.1(6.0)  &  1.0(9.1) 
\end{tabular}
\end{center}
\vspace{0mm}
\end{table}
\end{ruledtabular}

Due to the lack of knowledge about the scalar resonances 
at the simulated quark masses, 
we test a generic polynomial form up to the quartic order
\bea
   F_S(q^2)
   & = & 
   F_S(0) \left\{ 1 + a_{S,1}\, q^2 
                    + a_{S,2}\, (q^2)^2 
                    + a_{S,3}\, (q^2)^3 
                    + a_{S,4}\, (q^2)^4 \right\}
   \label{eqn:q2_interp:vs_q2:pff_s}
\eea
to parametrize the $q^2$ dependence of the scalar form factor $F_S(q^2)$.
We observe that the cubic ($a_{S,4}\!=\!0$) fit also describes our data 
reasonably well as seen in Fig.~\ref{fig:q2_interp:pff_s}. 
Fit results summarized in Table~\ref{tbl:q2_interp:vs_q2:pff_s} 
show that the inclusion of the quartic correction does not change 
the value of $\chi^2$ and 
the result for the scalar radius $\crad_S\!=\!6\, a_{S,1}$ significantly.
However, such a stability against the choice of the parametrization
is much less clear in the curvature $c_S\!=\!a_{S,2}$ 
due to its large uncertainty. 
From these observations,
we only use results for $\crad_S$ in the following analysis,
and leave a precise determination of $c_S$ for future studies.

%
%
%

%% file: 5.chiral_fit.tex
\section{Chiral extrapolation}
\label{sec:chiral_fit}

\subsection{Fit based on one-loop ChPT}


Since the form factors $F_{V,S}(q^2)$ are independent of $q^2$ at LO in ChPT, 
the chiral expansion of the radii $\crad_{V,S}$ starts from 
the one-loop order of ChPT.
We first compare our lattice results with the one-loop expressions
\cite{PFF:ChPT:NLO:1} 
\vspace{-1mm}
\bea
   \crad_V
   & = & 
   -\frac{1}{NF^2}\left( 1 + 6N\,l_6^r \right)
   -\frac{1}{NF^2}\ln\left[ \frac{M_\pi^2}{\mu^2} \right],
   \label{eqn:chiral_fit:r2_v:nlo}
   \\
   \crad_S
   & = &
    \frac{1}{NF^2}\left( -\frac{13}{2} + 6 N\,l_4^r \right)
   -\frac{6}{NF^2}\ln\left[ \frac{M_\pi^2}{\mu^2} \right],
   \label{eqn:chiral_fit:r2_s:nlo}
\eea
where $N\!=\!(4\pi)^2$, and
$F$ is the decay constant in the chiral limit.
We adopt the normalization of the decay constant 
$F_\pi\!=\!92$~MeV at the physical quark mass.
The renormalization scale $\mu$ is set to $4\pi F$ in our analysis.
At this order of the chiral expansion,
$F$ is the only LEC appearing in the $M_\pi$ dependent terms.
We fix this important parameter to $F\!=\!79.0(^{+5.0}_{-2.6})$~MeV,
which has been determined from our detailed analysis 
of the pion mass and decay constant \cite{Spectrum:Nf2:RG+Ovr:JLQCD}.
Each one of Eqs.~(\ref{eqn:chiral_fit:r2_v:nlo}) and 
(\ref{eqn:chiral_fit:r2_s:nlo}), therefore, has a single fit parameter, 
namely LECs $l_6^r$ or $l_4^r$ in their constant term.

\begin{figure}[t]
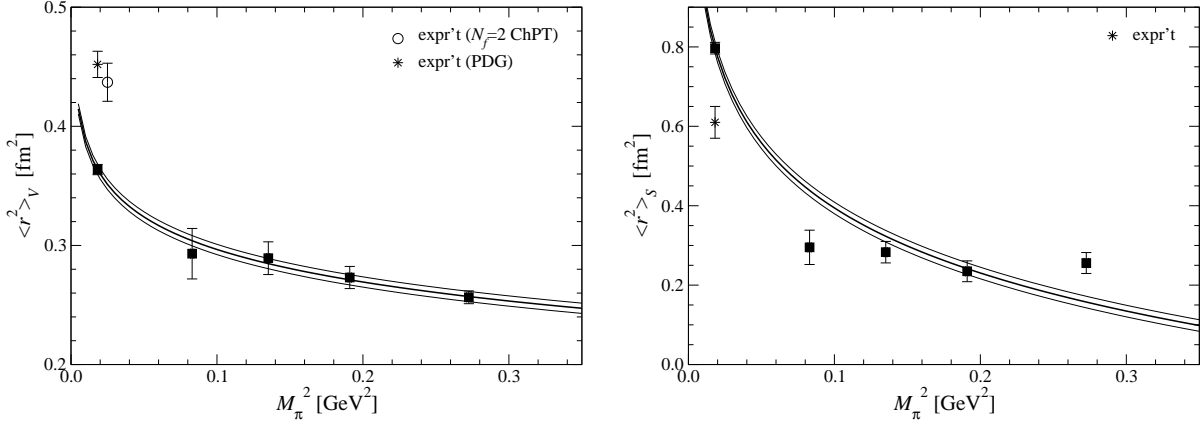

\begin{center}
\includegraphics[angle=0,width=0.48\linewidth,clip]{r2_v_vs_Mpi2.meas-pole+cubic.r2_v_nlo.eps}
\hspace{3mm}
\includegraphics[angle=0,width=0.48\linewidth,clip]{r2_s_vs_Mpi2.norm-mom0100.fh3.cubic.r2_s_nlo.eps}
\vspace{-3mm}
\caption{
   Chiral fit of $\crad_V$ (left panel) and $\crad_S$ (right panel)
   using one-loop ChPT formulae.
   Filled squares are the lattice data and the value extrapolated to the 
   physical point.
   In the left panel,
   we also plot the experimental value 
   $\crad_V\!=\!0.437(16)~\mbox{fm}^2$ 
   from an analysis based on $N_f\!=\!2$ ChPT \cite{PFF_V:ChPT:NNLO} 
   (open circle) 
   and $0.452(11)~\mbox{fm}^2$ quoted by Particle Data Group \cite{PDG:2008} 
   (star).
   The star symbol in the right panel represents 
   $\crad_S\!=\!0.61(4)~\mbox{fm}^2$ 
   obtained from an indirect determination 
   through $\pi\pi$ scattering \cite{PFF_S:ChPT:NNLO}.
}
\label{fig:chiral_fit:nlo}
\end{center}
\end{figure}

We find that 
the NLO fits of lattice data are not quite successful 
as seen in Fig.~\ref{fig:chiral_fit:nlo}
and Table~\ref{tbl:chiral_fit:r2:nlo}.
While the data of $\crad_V$ can be fitted 
with reasonable $\chi^2/{\rm d.o.f.}\!\sim\!0.14$,
the value extrapolated to the physical quark mass 
$\crad_V\!=\!0.3637(43)~\mbox{fm}^2$ 
is significantly smaller than the experimental value
$0.437(16)~\mbox{fm}^2$ based on $N_f\!=\!2$ ChPT \cite{PFF_V:ChPT:NNLO}
and $0.452(11)~\mbox{fm}^2$ quoted by PDG \cite{PDG:2008}.
As for the scalar radius,
the one-loop formula fails to reproduce our data of $\crad_S$ 
as indicated by the quite large value of $\chi^2/{\rm d.o.f.}\!\sim\!9$\,:
the data have a mild quark mass dependence 
in contrast to the $6$ times enhanced chiral logarithm 
compared to $\crad_V$.

\begin{ruledtabular}
\begin{table}[t]
\begin{center}
\caption{
   Results of chiral extrapolations of $\crad_V$ and $\crad_S$ 
   using one-loop ChPT formulae Eqs.~(\ref{eqn:chiral_fit:r2_v:nlo}) 
   and (\ref{eqn:chiral_fit:r2_s:nlo}).
   The radii extrapolated to the physical point are also listed.
}
\label{tbl:chiral_fit:r2:nlo}
\begin{tabular}{lll|lll}
   $\chi^2/{\rm d.o.f.}$  & $l_6^r \times 10^3$ & $\crad_V [\mbox{fm}^2]$ 
   \hspace{5mm} &
   $\chi^2/{\rm d.o.f.}$  & $l_4^r \times 10^3$ & $\crad_S [\mbox{fm}^2]$ 
   \\ \hline
   0.14 & $-$6.59(12) & 0.3637(43) &
   9.0  &    2.94(39) & 0.797(15)
   \\ 
\end{tabular}
\end{center}
\vspace{0mm}
\end{table}
\end{ruledtabular}

This failure of the NLO fits is not due to our choice of $F$.
If $F$ is treated as a free parameter, 
the fit to $\crad_S$ results in an unacceptably large value 
$F\!\simeq\!200$~MeV 
to achieve reasonable $\chi^2/{\rm d.o.f.} \lesssim 1$,
whereas
our data of $\crad_V$ favor $F\!\sim\!80$~MeV.
Thus we can not make a consistent analysis.
We note that 
we have experienced a similar situation,
namely $\crad_V$ smaller than experiment and 
a small quark mass dependence of $\crad_S$,
in our previous study with a different lattice action 
\cite{PFF:Nf2:Plq+Clv:JLQCD},
though simulated quark masses are heavier than those in this study.

We also note that the lattice data for the curvature $c_V$ largely 
deviate from its NLO ChPT expression as shown below.
From all of these observations we conclude that 
the chiral behavior of the pion form factors in the quark mass region 
$m_s/6$\,--\,$m_s/2$ are not well described by NLO ChPT.

\subsection{Fit based on two-loop ChPT}


In Ref.~\cite{Spectrum:Nf2:RG+Ovr:JLQCD},
we observed that 
NNLO contributions are important to reliably
extract LECs from the pion mass and decay constant 
in our simulated region of the quark mass.
Therefore there exists a possibility that NNLO contributions become
also important to describe the chiral behaviour of the radii.
%
%
%
The two-loop expressions of $\crad_{V,S}$ and $c_V$ 
are given by \cite{PFF_V:ChPT:NNLO,PFF_V:ChPT:NNLO:2}
\bea
   \crad_V
   & = & 
   -\frac{1}{NF^2}\left( 1 + 6N\,l_6^r \right)
   -\frac{1}{NF^2}\ln\left[ \frac{M_\pi^2}{\mu^2} \right]
   \nn \\
   &   &  
   +\frac{1}{N^2F^4}
    \left( \frac{13N}{192} - \frac{181}{48} + 6 N^2 r_{V,r}^r \right)\, M_\pi^2
   +\frac{1}{N^2F^4}
    \left( \frac{19}{6} - 12 N l_{1,2}^r \right)
    M_\pi^2 \ln\left[ \frac{M_\pi^2}{\mu^2} \right],
    \hspace{5mm}
   \label{eqn:chiral_fit:r2_v:nnlo}
\eea
\bea
   \crad_S
   & = & 
    \frac{1}{NF^2}\left( -\frac{13}{2} + 6 N\,l_4^r \right)
   -\frac{6}{NF^2}\ln\left[ \frac{M_\pi^2}{\mu^2} \right]
   \nn \\
   &   &
   +\frac{1}{N^2F^4}
    \left( - \frac{23N}{192} + \frac{869}{108} 
           + 88 N l_{1,2}^r + 80 N l_2^r + 5 N l_3^r
           - 24 N^2 l_3^r l_4^r
           + 6 N^2 r_{S,r}^r \right)\, M_\pi^2
   \nn \\
   & & 
   +\frac{1}{N^2F^4}
    \left( - \frac{323}{36} + 124 N l_{1,2}^r
                            + 130 N l_2^r \right)\, 
    M_\pi^2 \ln\left[ \frac{M_\pi^2}{\mu^2} \right] 
   -\frac{65}{3N^2F^4}
    M_\pi^2 \ln\left[ \frac{M_\pi^2}{\mu^2} \right]^2,
    \hspace{5mm}
   \label{eqn:chiral_fit:r2_s:nnlo}
   \\
   c_V
   & = & 
    \frac{1}{60NF^2} \frac{1}{M_\pi^2}
   +\frac{1}{N^2F^4}
    \left( \frac{N}{720} - \frac{8429}{25920} 
                         + \frac{N}{3}l_{1,2}^r
                         + \frac{N}{6}l_6^r
                         + N^2 r_{V,c}^r \right) 
   \nn \\
   &   & 
   +\frac{1}{N^2F^4}
    \left( \frac{1}{108} + \frac{N}{3} l_{1,2}^r
                         + \frac{N}{6} l_6^r \right)\, 
    \ln\left[ \frac{M_\pi^2}{\mu^2} \right] 
   +\frac{1}{72N^2F^4}
    \ln\left[ \frac{M_\pi^2}{\mu^2} \right]^2,
   \label{eqn:chiral_fit:c_v:nnlo}
\eea   
where we use a linear combination $l_{1,2}^r=l_1^r-l_2^r/2$
instead of $l_1^r$,
since the former is convenient for our chiral extrapolation (see below).
The analytic terms containing $r_{\{V,S\},\{r,c\}}^r$ represent 
contributions of tree diagrams with vertices 
from the $O(p^6)$ chiral Lagrangian.

\begin{ruledtabular}
\begin{table}[t]
\begin{center}
\caption{
   Phenomenological estimates of LECs used in this paper.
   We note that $O(p^6)$ couplings $r_{\{V,S\},\{r,c\}}^r$
   are based on a resonance saturation hypothesis.
}
\label{tbl:chiral_fit:nnlo:lec:pheno}
\begin{tabular}{l|lll|l|lllll}
   Ref.\cite{F:ChPT:NNLO} &
   \multicolumn{3}{c|}{Ref.\cite{PFF_S:ChPT:NNLO}} &
   Ref.\cite{PFF:ChPT:NLO:1} &
   \multicolumn{4}{c}{Ref.\cite{PFF_V:ChPT:NNLO}}
   \\ \hline   
   $F$[MeV]     & 
   $\bar{l}_1$  & $\bar{l}_2$  & $\bar{l}_4$  & 
   $\bar{l}_3$  & 
   $\bar{l}_6$  &
   $r_{V,r}^r \times 10^4$  & $r_{V,c}^r \times 10^4$  & 
   $r_{S,r}^r \times 10^4$  
   \\ \hline
   86.2(5)      & 
   $-$0.36(59)  & 4.31(11)     & 4.39(22)     & 
   2.9(2.4)     &
   16.0(9)      &
   $-$2.5       & 2.6          & $-$0.3         
   \\ 
\end{tabular}
\end{center}
\vspace{0mm}
\end{table}
\end{ruledtabular}

\begin{figure}[t]
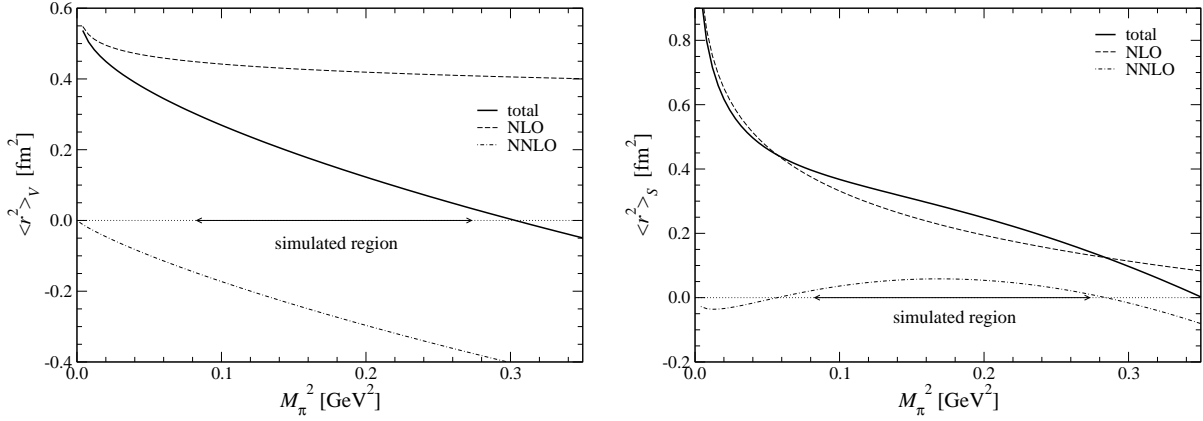

\begin{center}
\includegraphics[angle=0,width=0.48\linewidth,clip]{r2_v_vs_Mpi2_bct+cgl+cd.Mrho.eps}
\hspace{3mm}
\includegraphics[angle=0,width=0.48\linewidth,clip]{r2_s_vs_Mpi2_bct+cgl+cd.MS.eps}

\vspace{-3mm}
\caption{
   Chiral behavior of radii based on two-loop ChPT with 
   phenomenological estimates of LECs listed 
   in Table~\ref{tbl:chiral_fit:nnlo:lec:pheno}.
   The dashed and dot-dashed lines represent contributions at NLO and NNLO,
   whereas the solid line is their total.
   We note that 
   $r_{\{V,S\},\{r,c\}}^r$ from the resonance saturation are 
   taken as the renormalized LECs at the resonance mass scale in this plot.
}
\label{fig:chiral_fit:nnlo:bct+cgl+cd}
\vspace{0mm}
\end{center}
\end{figure}

Before fitting lattice data to these expressions,
one can get some idea about the significance of the NNLO contributions 
by using phenomenological estimates of LECs. 
A collection of recent estimates
is shown in Table~\ref{tbl:chiral_fit:nnlo:lec:pheno},
where 
the LECs in the $O(p^4)$ chiral Lagrangian are denoted by 
the scale invariant convention $\bar{l}_i$ defined by
\bea
   l_i^r
   = 
   \frac{\gamma_i}{2N}
   \left( \bar{l}_i + \ln\left[ \frac{M_\pi^2}{\mu^2} \right] \right)
   \label{eqn:chial_fit:lec_inv}
\eea
with 
\bea 
   \gamma_1 = \frac{1}{3}, \hspace{2mm}
   \gamma_2 = \frac{2}{3}, \hspace{2mm}
   \gamma_3 = -\frac{1}{2}, \hspace{2mm}
   \gamma_4 = 2, \hspace{2mm}
   \gamma_6 = -\frac{1}{3}.
   \label{eqn:chial_fit:gamma_i}
\eea
Figure~\ref{fig:chiral_fit:nnlo:bct+cgl+cd} shows the expected $M_\pi^2$
dependence of $\crad_V$ and $\crad_S$ from 
Eqs.~(\ref{eqn:chiral_fit:r2_v:nnlo}) and (\ref{eqn:chiral_fit:r2_s:nnlo})
with the phenomenological estimates of LECs.
The individual contributions from NLO and NNLO are also plotted.
This analysis suggests that 
the NNLO contributions could significantly modify 
the chiral behavior of the radii in our simulated quark masses.
However, we note that 
$r_{\{V,S\},\{r,c\}}^r$ are poorly known and 
these in Table~\ref{tbl:chiral_fit:nnlo:lec:pheno}
are determined from a resonance saturation hypothesis.
A chiral extrapolation of lattice data with the two-loop formulae
is therefore important 
to confirm the significance of the NNLO contributions
and to resolve the failure of the one-loop fit.

The curvature $c_V$ 
characterizes the $O(q^4)$ dependence of $F_V(q^2)$, 
and therefore requires the NNLO terms to describe its logarithmic dependence 
on the quark mass as well as a constant term.
Near the chiral limit it has a divergent term of the form 
$\sim 1/(F^2 M_\pi^2)$, which comes from non-analytic NLO contributions
in $F_V(q^2)$.
Since the divergent term $1/(F^2 M_\pi^2)$ is significant only 
below the physical pion mass, 
the analysis of the lattice data for $c_V$ requires the NNLO contributions.

\begin{figure}[t]
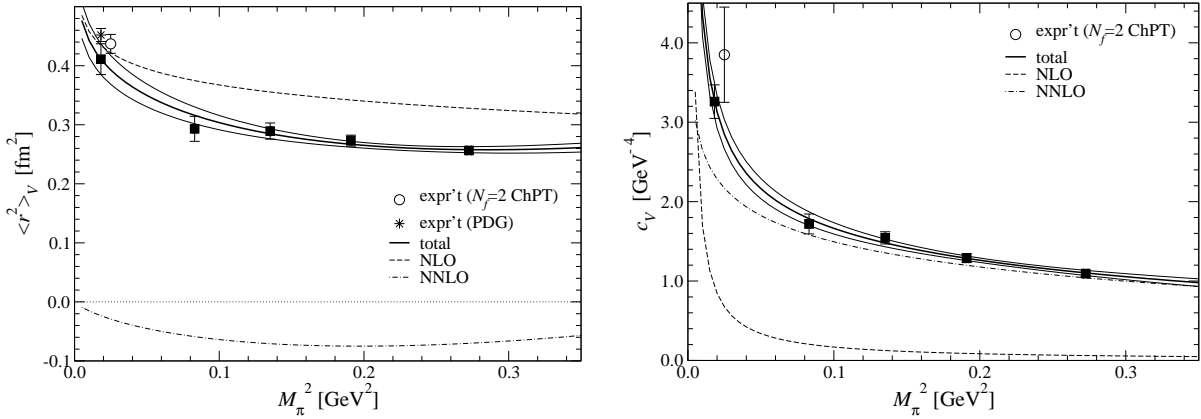

\begin{center} 
\includegraphics[angle=0,width=0.48\linewidth,clip]{r2_v_vs_Mpi2.meas-pole+cubic.r2_c_v.eps}
\hspace{3mm}
\includegraphics[angle=0,width=0.48\linewidth,clip]{c_v_vs_Mpi2.meas-pole+cubic.r2_c_v.eps}

\vspace{-3mm}
\caption{
   Simultaneous chiral fit to $\crad_V$ and $c_V$
   based on two-loop formulae
   Eqs.~(\ref{eqn:chiral_fit:r2_v:nnlo}) and (\ref{eqn:chiral_fit:c_v:nnlo}).
   The experimental value for $c_V\!=\!3.85(0.60)$ is taken from 
   Ref.~\cite{PFF_V:ChPT:NNLO}.
}
\label{fig:chiral_fit:nnlo:r2_c_v}
\vspace{-5mm}
\end{center}
\end{figure}

We extend our analysis to two-loop ChPT 
as already outlined in our previous report \cite{Lat08:JLQCD:TK}.
The curvature $c_V$ is included into our chiral extrapolation 
to obtain an additional constraint on LECs.
Both of $\crad_V$ and $c_V$ depend on 
$l_1^r$ and $l_2^r$ only through the linear combination $l_{1,2}^r$,
and 
the complicated two-loop expressions for $\crad_V$ and $c_V$ involve
only four free parameters $l_6^r$, $l_{1,2}^r$, $r_{V,r}^r$ and $r_{V,c}^r$
by choosing $M_\pi^2/(4 \pi F)^2$ as an expansion parameter.
Therefore
we first try a simultaneous fit to $\crad_V$ and $c_V$ 
to check that the chiral behavior of our data is described by two-loop ChPT.
Fit curves are plotted in Fig.~\ref{fig:chiral_fit:nnlo:r2_c_v} 
and numerical results are summarized in Table~\ref{tbl:chiral_fit:nnlo:r2_c_v}.
The fit leads to an acceptable value of $\chi^2/{\rm d.o.f.}\sim 0.7$,
and the relevant LECs are determined with a reasonable accuracy.
We note that this fit is based only on ChPT without additional assumptions.
The extrapolated values of $\crad_V$ and $c_V$ are consistent 
with recent phenomenological determinations 
using experimental data of $F_V(q^2)$ 
\cite{PFF_V:ChPT:NNLO,cV:DR:AR,cV:ChPT_DR:GHLM,r2+cV:Pade:MPS}.
%

\begin{ruledtabular}
\begin{table}[t]
\begin{center}
\caption{
   Results of simultaneous chiral fit to $\crad_V$ and $c_V$
   based on two-loop formulae
   Eqs.~(\ref{eqn:chiral_fit:r2_v:nnlo}) and (\ref{eqn:chiral_fit:c_v:nnlo}).
}
\label{tbl:chiral_fit:nnlo:r2_c_v}
\begin{tabular}{lllll|ll}
   $\chi^2/{\rm d.o.f.}$   & $l_6^r \times 10^3$     & ${l}_{12}^r \times 10^3$ 
                           & $r_{V,r}^r \times 10^5$ & $r_{V,c}^r \times 10^5$  
                           & $\crad_V [\mbox{fm}^2]$ 
                           & $c_V [\mbox{GeV}^{-4}]$ 
   \\ \hline
   0.70 & 
   $-$8.48(87) & $-$3.3(1.1)   & $-$0.77(69) & 3.97(20) &
   0.411(26)   & 3.26(21)
   \\ 
\end{tabular}
\end{center}
\vspace{0mm}
\end{table}
\end{ruledtabular}

In Sec.~\ref{sec:q2_interp}, 
we observe that the $O(q^6)$ contribution to $F_V(q^2)$ is not small
in the simulated region of the momentum transfer 
$q^2 \gtrsim 0.3~\mbox{GeV}^2$ ($q^2/(4 \pi F)^2 \gtrsim 0.3$).
The $q^2$ dependence of our data is therefore parametrized 
by the generic polynomial forms (\ref{eqn:q2_interp:vs_q2:pff_v})
and (\ref{eqn:q2_interp:vs_q2:pff_s}) instead of those based on ChPT.
This is not in contradiction with 
the successful chiral extrapolation of $\crad_V$ and $c_V$:
since we explore small pion masses 
$M_\pi^2 \lesssim 0.3~\mbox{GeV}^2$
($M_\pi^2/(4 \pi F)^2 \lesssim 0.3$),
the quark mass dependence of $\crad_V$ and $c_V$ is described by the 
two-loop ChPT formulae.

\begin{figure}[t]
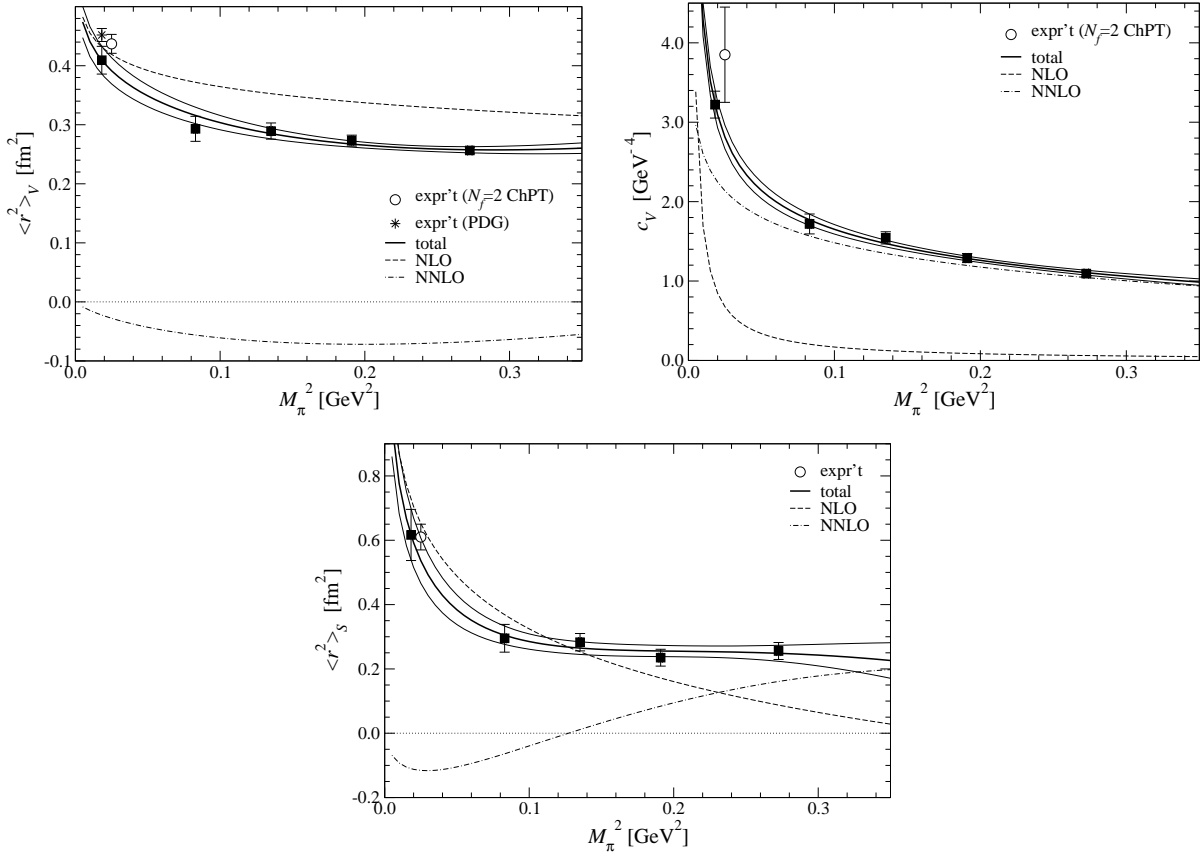

\begin{center}
\includegraphics[angle=0,width=0.48\linewidth,clip]{r2_v_vs_Mpi2.meas-pole+cubic.fh3-cubic.r2_v_s_c_v.vs_Mpi2.eps}
\hspace{3mm}
\includegraphics[angle=0,width=0.48\linewidth,clip]{c_v_vs_Mpi2.meas-pole+cubic.fh3-cubic.r2_v_s_c_v.vs_Mpi2.eps}
\vspace{3mm}

\includegraphics[angle=0,width=0.48\linewidth,clip]{r2_s_vs_Mpi2.meas-pole+cubic.fh3-cubic.r2_v_s_c_v.vs_Mpi2.eps}
\vspace{-3mm}
\caption{
   Simultaneous chiral fit for radii $\crad_{V,S}$ and curvature $c_V$.
}
\label{fig:chiral_fit:nnlo:r2_v_s_c_v}
\vspace{-5mm}
\end{center}
\end{figure}

The inclusion of $\crad_S$ into the simultaneous chiral fit introduces 
additional four free parameters  $l_2^r$, $l_3^r$, $l_4^r$ and $r_{S,r}^r$,
and we need to fix some of them to obtain a stable fit.
Since $l_2^r$ and $l_3^r$ appear only in the NNLO terms
and have been determined with a reasonable accuracy 
from phenomenology or lattice studies,
we use a phenomenological estimate $\bar{l}_2\!=\!4.31(11)$ 
\cite{PFF_S:ChPT:NNLO} and 
a lattice estimate $\bar{l}_3\!=\!3.38(56)$ from our analysis 
of the pion spectroscopy \cite{Spectrum:Nf2:RG+Ovr:JLQCD}.
We treat $r_{S,r}^r$ and $l_4^r$ as free parameters
because of poor knowledge on the former 
and in order to examine the consistency of the latter
with that determined from $F_\pi$.
The fit curves are shown in Fig.~\ref{fig:chiral_fit:nnlo:r2_v_s_c_v}
and numerical results are summarized 
in Table~\ref{tbl:chiral_fit:nnlo:r2_v_s_c_v}.
We observe that
i) the results in the vector channel,
namely  $\crad_V$, $c_V$ and the relevant LECs, 
do not change significantly by including $\crad_S$ into the chiral fit,
and ii) 
this fit describes the mild quark mass dependence of $\crad_S$ 
with an acceptable value of  $\chi^2/{\rm d.o.f.}\!\sim\!0.7$.
From Fig.~\ref{fig:chiral_fit:nnlo:r2_v_s_c_v},
we find that
the net NNLO contribution to $\crad_S$ is larger than NLO around 
our largest quark mass.
This is due to an accidental cancellation between 
the constant and logarithmic terms at NLO.

\begin{ruledtabular}
\begin{table}[t]
\begin{center}
\caption{
   Results of simultaneous chiral fit to $\crad_{V,S}$ and $c_V$
   based on two-loop formulae
   Eqs.~(\ref{eqn:chiral_fit:r2_v:nnlo}), (\ref{eqn:chiral_fit:r2_s:nnlo})
   and (\ref{eqn:chiral_fit:c_v:nnlo}).
}
\label{tbl:chiral_fit:nnlo:r2_v_s_c_v}
\begin{tabular}{lllllll|llll}
   $\chi^2/{\rm d.o.f.}$   & $l_6^r \times 10^3$     & $l_4^r \times 10^3$  
                           & $l_{1,2}^r \times 10^3$  
                           & $r_{V,r}^r \times 10^5$  
                           & $r_{V,c}^r \times 10^5$  
                           & $r_{S,r}^r \times 10^4$  
                           & $\crad_V [\mbox{fm}^2]$ 
                           & $c_V [\mbox{GeV}^{-4}]$ 
                           & $\crad_S [\mbox{fm}^2]$ 
   \\ \hline
   0.68 & $-$8.41(76)  & 1.1(3.2)  & $-$3.10(90) 
        & $-$1.0(1.1)  & 4.00(17)  & 1.74(36)
        & 0.409(23)    & 3.22(17)  & 0.617(79)
   \\ 
\end{tabular}
\end{center}
\vspace{0mm}
\end{table}
\end{ruledtabular}

We take results from this simultaneous fit 
as our best estimate of $\crad_{V,S}$, $c_V$. 
The systematic error due to the chiral extrapolation is estimated 
by excluding the data at the largest quark mass from the fit.
The shifts in the radii and curvature turn out to be at the 1\,$\sigma$ level: 
namely, the extrapolation is stable against variation of the fit range.
In order to estimate systematics due to 
the fixed topology and the use of the Feynman-Hellmann theorem to 
estimate $F_S(0)$, 
we repeat the whole analysis by using $F_{V,S}(q^2)$ 
shifted by their systematic uncertainties discussed
in Secs.~\ref{sec:pff:pff_s} and \ref{sec:pff:fixedQ}.
Systematic uncertainties 
due to the choice of the inputs,
namely LECs $F$, $l_2^r$, $l_3^r$ for the chiral extrapolation and 
$r_0\!=\!0.49$~fm to fix the scale, 
are estimated by shifting each LEC by its uncertainty 
and by using a recent lattice estimate $r_0\!=\!0.47$~fm \cite{r0:Nf2:ImpG+Asqt:MILC:1,r0:Nf2:ImpG+Asqt:MILC:2}.
In addition, 
a discretization error estimated by a naive order counting
$O((a\Lambda_{\rm QCD})^2) \approx 3$\,\% is also taken into account.
The magnitude of these systematic uncertainties are summarized in 
Table~\ref{tbl:chiral_fit:nnlo:error}.
Our final results for the radii and curvature at the physical point are 
\bea
   \crad_V   & = & 0.409(23)(37)~\mbox{fm}^2,
   \\
   \crad_S   & = & 0.617(79)(66)~\mbox{fm}^2,
   \\
   c_V       & = & 3.22(17)(36)~\mbox{GeV}^{-4},
\eea 
where the first error is statistical and 
the second represents systematic errors added in quadrature.
These results are in good agreement with their phenomenological values. 
The largest systematic error arises from the use of the input $r_0$
to fix the scale.
This can be removed by a better choice, such as $f_\pi$,
in future studies in three-flavor QCD.

\begin{ruledtabular}
\begin{table}[t]
\begin{center}
\caption{
   Statistical and systematic errors of radii and curvature.
   We list the magnitude of the errors relative to the central values.
}
\label{tbl:chiral_fit:nnlo:error}
\begin{tabular}{l|ll|lllllll}
             & stat. & sys.(total)
             & chiral fit & input(LECs) & input($r_0$) 
             & fixed $Q$  & Feynman-Hellmann   
             & finite $a$ 
   \\
   $\crad_V$ &  6\,\%   &  9\,\%
             &  5.3\,\% & 0.2\,\% & 6.1\,\% & 2.2\,\% & 1.4\,\% & 3.0\,\% 
   \\
   $\crad_S$ &  13\%    & 11\%
             &  4.0\,\% & 2.2\,\% & 7.7\,\% & 2.4\,\% & 4.6\,\% & 3.0\,\%
   \\ 
   $c_V$     &  5\,\%   & 11\,\%
             &  2.7\,\% & 0.4\,\% & 9.3\,\% & 2.3\,\% & 3.8\,\% & 3.0\,\% 
   \\ 
\end{tabular}
\end{center}
\vspace{0mm}
\end{table}
\end{ruledtabular}

We estimate the relevant LECs in a similar way to that for 
$\crad_{V,S}$ and $c_V$, and obtain 
\bea 
   \bar{l}_6 & = & 11.9(0.7)(1.0),
   \\
   \bar{l}_4 & = & 4.09(50)(52),
   \\
   \bar{l}_1-\bar{l}_2 & = & -2.9(0.9)(1.3),
   \label{eqn:chiral_fit:result:l12}
   \\
   r_{V,r}^r & = & -1.0(1.0)(2.5) \times 10^{-5},
   \\
   r_{V,c}^r & = & 4.00(17)(64) \times 10^{-5},
   \\
   r_{S,r}^r & = & 1.74(36)(78) \times 10^{-4}.
\eea   
%
Our estimate of $\bar{l}_6$ is slightly smaller than 
those from two-loop ChPT analyses of experimental data;
$\bar{l}_6\!=\!16.0(0.9)$ from $F_V$ \cite{PFF_V:ChPT:NNLO}
and 15.2(0.4) from $\tau$ and $\pi$ decays \cite{l6:ChPT:NNLO:tau+pi}.
This is partly due to the deviation of $F$ 
between our lattice determination \cite{Spectrum:Nf2:RG+Ovr:JLQCD} 
and two-loop ChPT \cite{F:ChPT:NNLO}.
We note that $\bar{l}_4$ is consistent with our lattice determination
$\bar{l}_4\!=\!4.12(56)$ from $F_\pi$ \cite{Spectrum:Nf2:RG+Ovr:JLQCD}
and a phenomenological estimate 4.39(22) \cite{PFF_S:ChPT:NNLO}.
Here we list $\bar{l}_1\!-\!\bar{l}_2$ instead of $l_{1,2}^r$,
which is scale invariant and 
equals to $(\bar{l}_1\!-\!\bar{l}_2)/(6N)$
in the convention of Eq.~(\ref{eqn:chial_fit:lec_inv}).
Although $\bar{l}_1\!-\!\bar{l}_2$ has a large uncertainty of 
$\sim$ 50\,\%, it is consistent with $-4.67(60)$ from phenomenology.
Note that 
we set the renormalization scale $\mu\!=\!4 \pi F$, 
and our fit favors small values of the order of $10^{-4}$\,--\,$10^{-5}$
for the renormalized $O(p^6)$ couplings $r^r_{\{V,S\},\{r,c\}}$ at this scale.
Finally, we emphasize that 
the simultaneous fit only to $\crad_V$ and $c_V$ 
without any phenomenological input 
leads to consistent results for the vector channel,
namely for 
$\crad_V$, $c_V$, $\bar{l}_6$, $\bar{l}_1\!-\!\bar{l}_2$ and $r_{V,\{r,c\}}^r$.

%% file: 6.conclusion.tex

\section{Conclusions} 
\label{sec:concl}


In this article,
we present a lattice calculation of the pion form factors $F_{V,S}(q^2)$ 
in two-flavor QCD.
We study the chiral behavior of the radii $\crad_{V,S}$ and 
the curvature $c_V$ based on ChPT up to two loops.
We employ the overlap quark action,
which has exact chiral symmetry 
and thus provides the cleanest framework for the study of chiral behavior.
Otherwise, 
the explicit chiral symmetry breaking of conventional lattice actions 
could make studies based on two-loop ChPT substantially complicated.
Another salient feature of this work is that, for the first time,
$F_S(q^2)$ is evaluated including the contributions of the disconnected 
diagrams through the all-to-all quark propagators.


Our detailed analysis based on ChPT reveals that 
two-loop contributions are important 
to describe the chiral behavior of $\crad_{V,S}$ and $c_V$
at our simulated quark masses, 
which are comparable to those in recent unquenched simulations.
Through the chiral extrapolation at two loops,
we obtain $\crad_{V,S}$ and $c_V$ at the physical point,
which are consistent with experiment. 
We also obtain estimates of $O(p^4)$ and $O(p^6)$ LECs,
and confirm that 
$F_S(q^2)$ and $F_\pi$ lead to consistent results for $\bar{l}_4$
as suggested long ago \cite{PFF:ChPT:NLO:1}.

 
%
%
As we already outlined in Ref.~\cite{Lat08:JLQCD:TK},
the curvature $c_V$ is useful to stabilize the two-loop chiral fit.
The single pole ansatz Eq.~(\ref{eqn:q2_interp:vs_q2:pff_v:pole}),
which has been commonly used in previous studies,
may not be suitable to estimate $c_V$,
since it simply assumes a relation $c_V\!=\!(\crad_V/6)^2$
at simulated quark masses.
In this work, instead,
we employ a generic form up to cubic corrections 
to parametrize the $q^2$ dependence of $F_{V,S}(q^2)$
thanks to the precise determination of $F_{V,S}(q^2)$ 
through the all-to-all propagator.
Dispersive analyses of the $q^2$ dependence, 
model independent information of scalar resonances,
and the twisted boundary condition \cite{TBC} 
already used in Refs.\cite{PFF:Nf3:RG+DW:RBC+UKQCD,PFF:Nf2:Sym+tmW:ETMC}
are interesting subject and technique 
for a better control of this parametrization in future studies.
%

%
%
Extending this study to three-flavor QCD
is important for more realistic comparison with experiment.
We have already started simulations in three-flavor QCD 
with the overlap action \cite{Lat07:JLQCD:Matsufuru,Lat08:JLQCD:Hashimoto};
measurements of pion correlators using all-to-all propagators 
are in progress.

%
%
Finally, 
it is expected from our chiral extrapolation that 
$\crad_S$ shows a strong quark mass dependence due to 
the one-loop chiral logarithm below $M_\pi \! \sim \! 250$~MeV. 
Pushing simulations toward such small quark masses 
is an interesting subject in future studies
for a direct observation of the one-loop logarithm,
although it is very challenging.


\begin{acknowledgments}

Numerical simulations are performed on Hitachi SR11000 and 
IBM System Blue Gene Solution 
at High Energy Accelerator Research Organization (KEK) 
under a support of its Large Scale Simulation Program (No.~08-05).
This work is supported in part by the Grant-in-Aid of the
Ministry of Education (No.~18340075, 18740167, 19540286, 19740160,
20025010, 20039005, 20105001, 20105002, 20105003, 20105005, 20340047, 
20740156 and 21684013),
the National Science Council of Taiwan
(No.~NSC96-2112-M-002-020-MY3, NSC96-2112-M-001-017-MY3, NSC97-2119-M-002-001), 
and NTU-CQSE (No.~97R0066-65 and 97R0066-69).

\end{acknowledgments}